\documentclass{aa}  

\usepackage{graphicx}
\usepackage{derivative}
\usepackage{xcolor}
\usepackage{float}
\usepackage{rotating}
\usepackage{booktabs}
\usepackage{mathtools}
\usepackage{placeins}
\usepackage{bm}
\usepackage{longtable}
\usepackage{adjustbox}
\usepackage{supertabular}
\usepackage{multirow}
\usepackage{tabularx}
\usepackage{url}
\usepackage{esdiff}
\usepackage{amsmath}
\usepackage{siunitx}

\defcitealias{Berthomieu1990}{BP90}
\defcitealias{Breton2022simuFstars}{B22a}
\defcitealias{Huber2013}{H13}
\defcitealias{Mathur2014}{M14}
\defcitealias{Lund2017}{L17}

\usepackage{hyperref}
\hypersetup{
  colorlinks=true,
    linkcolor=blue,
    citecolor=blue,
    urlcolor=magenta}

\usepackage{natbib}

\usepackage{txfonts}

\begin{document}

   \title{In search of gravity mode signatures \\ in main sequence solar-type stars observed by \textit{Kepler}}
    
   \titlerunning{}

   \author{S.N.~Breton\inst{1,2}
          \and
          H. Dhouib\inst{2}
          \and
          R.A.~Garc\'{i}a\inst{3}
          \and
          A.S~Brun\inst{3}
          \and
          S.~Mathis\inst{3}
          \and
          F.~Pérez Hernández\inst{4,5}
          \and
          S.~Mathur\inst{4,5}
          \and
          A.~Dyrek\inst{2}
          \and
          A.R.G.~Santos\inst{6}
          \and 
          P.L.~Pallé\inst{4,5}
          }
    \institute{INAF – Osservatorio Astrofisico di Catania, Via S. Sofia, 78, 95123 Catania, Italy
    \and
    Universit\'e Paris Cit\'e, Universit\'e Paris-Saclay, CEA, CNRS, AIM, F-91191, Gif-sur-Yvette, France \\
    \email{sylvain.breton@inaf.it}
    \and
    Universit\'e Paris-Saclay, Universit\'e Paris Cit\'e, CEA, CNRS, AIM, F-91191, Gif-sur-Yvette, France 
    \and Instituto de Astrof\'{i}sica de Canarias, 38205 La Laguna, Tenerife, Spain
    \and Departamento de Astrof\'{i}sica, Universidad de La Laguna (ULL), 38206 La Laguna, Tenerife, Spain 
    \and Instituto de Astrof\'isica e Ci\^encias do Espa\c{c}o, Universidade do Porto, CAUP, Rua das Estrelas, PT4150-762 Porto, Portugal
    }

   \date{}

 \abstract{
 Gravity modes (g modes), mixed gravito-acoustic modes (mixed modes), and gravito-inertial modes (gi modes) possess unmatched properties as probes for stars with radiative interiors. The structural and dynamical constraints that they are able to provide cannot be accessed by other means. While they provide precious insights into the internal dynamics of evolved stars as well as massive and intermediate-mass stars, their non-detection in main sequence (MS) solar-type stars make them a crucial missing piece in our understanding of angular momentum transport in radiative zones and stellar rotational evolution. 
 In this work, we aim to apply certain analysis tools originally developed for helioseismology in order to look for g-mode signatures in MS solar-type stars. 
 We select a sample of the 34 most promising MS solar-type stars with \textit{Kepler} four-year long photometric time series. All these stars are well-characterised late F-type stars with thin convective envelopes, fast convective flows, and stochastically excited acoustic modes (p~modes). For each star, we compute the background noise level of the Fourier power spectrum to identify significant peaks at low frequency. 
After successfully detecting individual peaks in 12 targets, we further analyse four of them and observe distinct patterns of surrounding peaks with a low probability of being noise artifacts. Comparisons with the predictions from reference models suggest that these patterns are compatible with the presence of non-asymptotic low-order pure g modes, pure p modes, and mixed modes. Given their sensitivity to both the convective core interface stratification and the coupling between p- and g-mode resonant cavities, such modes are able to provide strong constraints on the structure and evolutionary states of the related targets. Considering the granulation and activity background of the stars in our sample, we subsequently compute the corresponding mode velocity necessary to trigger a detectable luminosity fluctuation. We use it to estimate the surface velocity, $\left<v_r\right>$, of the candidate modes we have detected. In this case, we find $\left<v_r\right> \sim 10$~cm/s. These results could be extremely useful for characterising the deep interior of MS solar-type stars, as the upcoming PLATO mission will considerably expand the size of the available working sample. 
 }


   \maketitle

\section{Introduction \label{section:introduction}}

Manifestations of gravity modes (g modes) are observed in numerous regions of the Hertzsprung-Russell (HR) diagram. In main sequence (MS) massive stars \citep[e.g.][]{Waelkens1991} and intermediate-mass stars \citep[e.g.][]{Kaye1999}, they are most often detected in the form of gravito-inertial modes (gi-modes) as these objects are fast rotators with large Coriolis frequencies (i.e. twice the rotation frequency, a quantity that characterises the importance of rotation for the dynamics of waves). 
In particular, these gi-modes allowed mixing \citep[e.g.][]{Degroote2010,Pedersen2022} and rotation \citep[e.g.][]{VanReeth2015,VanReeth2016,VanReeth2018,Ouazzani2017,Li2020} to be probed at the convective core boundary.
At more evolved stages, in subgiant \citep[e.g.][]{Kjeldsen1995} and red giant \citep[e.g.][]{Beck2011} stars, these modes are observed as mixed gravito-acoustic modes with coupled resonant cavities, providing information on the evolutionary stage \citep[e.g.][]{Bedding2011}, the core-to-surface \citep[e.g.][]{Beck2012,Deheuvels2015} and core \citep[e.g.][]{Mosser2012,Gehan2018} rotation, and on the deep-layer magnetism of the star \citep{Li2022,Deheuvels2023}. Finally, after the loss of the envelope, g modes are detected in white dwarfs \citep[e.g.][]{Winget1991}. 
However, on the MS, below the $\gamma$~Dor instability strip \citep[e.g.][]{Guzik2000}, no star has yet  been found to exhibit an observable g-mode signature the exception being the Sun itself, with detection claims presented, for example, by \citet{Garcia2007} and \citet{Fossat2017}\footnote{For an extensive review and discussion of these two claims and the history of the search of solar g modes, we refer the reader to \citet{Appourchaux2013}, Sect.~1 from \citet{Breton2022}, and \citet{Belkacem2022}.}. 
This is due to the large convective envelope surrounding the radiative zone in such stars, a medium where internal gravity waves  (IGWs) are evanescent because of the imaginary character of the Brunt-Väisälä frequency in convective layers \citep[e.g.][]{Kippenhahn2012,ChristensenDaalsgardLectureNotes}.
The task of detecting these low-amplitude surface-evanescent modes is made even more difficult by the fact that the dominant signal in the frequency intervals where g modes are expected results from convective granulation \citep[e.g.][]{Harvey1985,Mathur2011,Kallinger2014}. In order to simultaneously characterise the non-linear interplay between IGWs and convection, the g-mode power spectrum, and the mode surface visibility, 2D and 3D deep-layer numerical simulations of the Sun have been performed \citep[e.g.][]{Brun2011,Alvan2014,Alvan2015,LeSaux2022}.

Stellar internal rotation can be seismically characterised by measuring the rotational splittings between $m$-components of a given mode of radial order $n$ and degree $\ell$, as rotation lifts the frequency degeneracy between azimuthal number $m$ \citep[e.g.][]{ChristensenDaalsgardLectureNotes}. Specifically with gi modes, another way is to measure the evolution with frequency of the period spacing $\Delta P_{\ell,n}$, that is, the period difference between the two modes of the same $\ell$ and $m$, but with consecutive radial orders $n$ and $n+1$.
The characterised properties of gi-modes (for intermediate-mass stars) and mixed modes (for low- and intermediate-mass evolved stars) suggest radiative zone rotation profiles compatible with weak radial differential rotation \citep[see e.g.][and references therein]{Aerts2019}.
Evidence of the same behaviour for solar-type stars was provided by the pressure-mode (p-mode) analysis performed by \citet{Benomar2015}.
Therefore, detecting and characterising gi-mode $\Delta P_\ell$ or g, gi, and/or mixed rotational splittings in MS solar-type pulsators is of utmost importance in order to put constraints on the structure and the rotation profile of their innermost internal regions. 

Late F-type stars have thin convective envelopes and exhibit p-modes stochastically excited by turbulent motions in the upper layer of the convective envelope. Stochastic mechanisms should also excite g~modes in these stars via convective pummeling at the radiative--convective zone interface \citep[e.g.][]{Press1981,Pincon2016} and via bulk turbulence at the bottom of the convective envelope \citep[e.g.][]{Goldreich1990,Belkacem2009}. 
Convective velocity in the envelope of solar-type stars is expected to scale as the cube of stellar mass \citep[e.g.][]{Brun2017a}.
As pointed out by the 3D deep-shell hydrodynamical simulations of rotating 1.3 $M_\odot$ stars performed by \citet[][hereafter B22a]{Breton2022simuFstars}, faster convective flows and reduced tunnelling distance should therefore both favour a significantly increased surface g-mode amplitude compared to the solar case \citep[e.g.][]{Shibahashi1979,Mathis2014,Pincon2016,Mathis2023Tunneling}. Late F-type stars could therefore represent our best candidates to accurately characterise the deep radiative interior rotation of MS solar-type stars, as well as their internal mixing and deep magnetism. In addition, these stars have recently attracted the interest of the community in relation to their magnetic activity properties \citep{Mathur2014}, the specific challenge they represent concerning the modelling of angular momentum transport along their life on the MS \citep{Betrisey2023}, and the penetrative convection at the top of the radiative interior \citep{Deal2023}.

The practical, statistical, and theoretical tools developed over the years to study g-mode properties in solar-type stars have   been, until now, almost exclusively applied to the solar case. Having described the properties that would favour strong stochastic excitation of g modes in higher-mass solar-type stars, we therefore propose the application of some of these tools to targets observed almost continuously over four years by the \textit{Kepler} satellite \citep{Borucki2010} and exhibiting the stochastic variability related with solar-type p-mode oscillations. Approaching the lower boundary of the $\gamma$~Dor instability strip in the HR diagram, the detection of stochastic p modes is crucial for us to select late F-type stars that still have solar-type properties while having the larger convective velocities and the thinner envelope. Indeed, the mode stochasticity constitutes undeniable evidence that the star possesses a convective envelope with characteristics similar to the solar convection zone. Moreover, through modelling works \citep[e.g.][]{Deheuvels2016,SilvaAguirre2017,Creevey2017}, p modes provide the best available insights concerning the internal structure of the stars we want to consider. 
Fortunately, the \textit{Kepler} short-cadence mode allowed the characterisation of solar-type p modes in several hundred MS stars \citep[e.g.][]{Chaplin2011,Mathur2022}. 

The layout of the paper is as follows. 
In Sect.~\ref{sec:sample_selection}, we present the selection rules we followed to obtain our working sample and describe the important properties of the stars that compose it.
In Sect.~\ref{sec:ensemble_analysis}, we describe the method we used to obtain an approximation of the background profile and we present a homogeneous statistical analysis that we performed in order to search for significant individual peaks. 
Using the corresponding detections, we selected peak patterns with a low probability of being  induced by noise. By comparing the observations with predictions from reference models computed for this work, we identified four stars of interest. In Sect.~\ref{sec:kic3733735}, we discuss the case of KIC~3733735, for which we emphasise that the detected pattern is compatible with a series of non-asymptotic pure g modes followed by the $n=1$ pure p modes. In Sect.~\ref{sec:mixed_modes}, we present a study of the cases of KIC~6679371, KIC~7103006, and KIC~9206432, where the models suggest we should witness a coupling between cavities, which should result in decreased inertia for the g-dominated mixed modes.
 In Sect.~\ref{sec:upper_threshold}, we connect the luminosity fluctuations observed in \textit{Kepler} to maximal radial displacement and velocity at the surface of the considered stars for modes of different frequencies and degrees.  
We draw conclusions from this work and suggest some perspectives in Sect.~\ref{section:conclusion}. 

\section{Sample selection \label{sec:sample_selection}}

Our targets of interest were selected among the stars observed in short cadence by \textit{Kepler} and exhibiting stochastically excited p modes.  Table~\ref{tab:fstar_selection} summarises the global properties of the selected targets. 
When available, we consider effective temperature $T_\mathrm{eff}$, logarithm of the surface gravity, $\log g$, mass $M_\star$, and metallicity index $[\rm Fe/H]$ from \citet{SilvaAguirre2017}, using the outputs from the ASTFIT modelling. Otherwise, we consider values from the \textit{Kepler} Data Release 25 \citep[DR25][]{Mathur2017}.
We therefore considered stars from \citet{Mathur2014}, \citet{Lund2017}, and \citet{Hall2021}, with $6200 < T_\mathrm{eff} < 6900$~K and $\log g > 3.9$.
No additional target was recovered by inspecting the \citet{Mathur2022} sample, which is so far the most complete catalogue of \textit{Kepler} MS solar-type pulsators.
We note that KIC~9226926 is the only member of our sample with $T_\mathrm{eff} = 6887$~K above 6700~K, which positions it very close to the lower edge of the $\gamma$-Dor instability strip. However, KIC~9226926 is clearly a solar-type star, with signatures of spot-activity modulations and stochastic p modes in its Fourier spectrum. 
For these reasons, we tend to think that its $T_\mathrm{eff}$ value available in DR25, albeit consistent with the $T_\mathrm{eff}=6853$~K value provided in the \textit{Gaia-Kepler} catalogue from \citet{Berger2020}, might be overestimated. We note that \citet{Mathur2014} used $T_\mathrm{eff} = 7149$~K, a value that is even higher.
We finally underline that the $T_\mathrm{eff}$, $\log g$, $M_\star$, and $[\rm Fe/H]$ values we compiled for this work are mainly indicative, and that their accuracy barely affects our subsequent analysis, which is focused on the stellar light curves acquired by \textit{Kepler}. Combining fast convective flows and relatively thin convective envelope, this is the most promising sample that can possibly be built from \textit{Kepler} observations in order to attempt stochastic g-mode detections in MS solar-type stars.  

The light curves that we use\footnote{The light curves can be downloaded from the following repository: \url{gitlab.com/sybreton/f-type-stars-gravity-modes}.} in what follows are calibrated with an improved version of the KEPSEISMIC method \citep{Garcia2011,Garcia2014,Pires2015} and filtered with a high-pass filter with a cutoff at 55 days;
they differ from the KEPSEISMIC light curves previously made accessible in the Mikulski Archive from Space Telescopes (MAST) in the sense that they are produced from a more recent version of the raw data available on MAST\footnote{The pixel files were downloaded from \url{archive.stsci.edu/kepler/data_search/search.php} in May 2022 and correspond to \textit{Kepler} DR25.}, and that an additional correction step to reduce the effect of the \textit{Kepler} annual modulation was carried out prior to the filtering. This additional correction step is the following: the auto-correlation function of the light curve is computed and the maximum in the interval $372.5 \pm 30$ days is identified, with 372.5 days corresponding to the period of the \textit{Kepler} orbit. The light curve is phase-folded over the corresponding period and a wavelet filtering is applied \citep{Starck2010}. The light curves produced with this new procedure are significantly less perturbed by \textit{Kepler} orbital modulations.

To analyse frequency regions below 283.2~$\mu$Hz (i.e. the Nyquist frequency,  $\nu_N$, of the long cadence sampling of \textit{Kepler}), we prefer the long-cadence data over the short-cadence data considering their longer extent in time. Indeed, \textit{Kepler} quasi-continuous short-cadence light curves are only available from \textit{Kepler} Quarter 5 onwards. 
The long-cadence sampling may result in a loss of power in the stellar signal close to the Nyquist frequency \citep[e.g.][]{Chaplin2011ApJ,Kallinger2014} that is proportionally to a factor $\eta^2$ given by
\begin{equation}
    \eta^2 = \mathrm{sinc}^2 \left( \frac{\pi \nu}{2 \nu_N} \right) \; ,
\end{equation}
where $\nu$ is the frequency.
For $\nu_N = 283.2$~$\mu$Hz, this corresponds to a loss in the power spectral density (PSD) of about 10\% at $\nu = 100$~$\mu$Hz, and about 35\% at $\nu = 200$~$\mu$Hz, which may have consequences in terms of the detectability of low-order (high-frequency) g modes. In amplitude, this corresponds to losses of $\sim$5\% and $\sim$19\%, respectively. Nevertheless, we consider this to be an acceptable trade-off given the significant extension of resolution offered by the long-cadence time series.
For our sample, the frequency of maximal oscillation amplitude, $\nu_\mathrm{max}$, is significantly beyond the Nyquist frequency of the long-cadence observations. Their contribution is therefore averaged out during signal integration, and we do not expect any aliasing related to the high-order p modes in the PSD, as can be observed for red giants with $\nu_\mathrm{max} \sim \nu_N$ \citep{Chaplin2014b}.

We note that one of the stars selected with the rules stated above is the exoplanet-host star HAT-P7a \citep[KIC~10666592, see e.g.][]{Pal2008}. We decided to exclude it from our sample, because even though our calibration removes the major contribution of the transit events, the HAT-P7b planetary-induced modulation \citep[see e.g.][]{Shporer2017} broadly contaminates the low-frequency region of the PSD for this target. 
KIC~3425851, KIC~4349452, KIC~7670943, KIC~8292840, KIC~8866102, and KIC~11807274 are also planet-host stars \citep[e.g.][]{Davies2016}. 
KIC~886102 is also known to have a K-type companion in its vicinity \citep{VanEylen2014}.
However, we keep these stars in our sample as we do not find evidence of PSD contamination by the planetary out-of-transit light modulation.
It should be noted that, from \textit{Gaia} \citep{Gaia2016} astrometric measurements \citep{Holl2023} and \textit{Kepler} seismic rotational splittings \citep{Hall2021}, KIC~7206837 is a candidate synchronised binary \citep{Ball2023}, as is KIC~10355856.
Finally, the light curves of two further targets also exhibit stable modulations with a period of about one day. These modulations cannot be related to spot activity and are evidence of the presence of a close-orbiting non-transiting companion. We also removed them from our working sample as the tidal interactions leading to possible g-mode excitation and resonance are complex \citep[see e.g.][]{Fuller2017,Bunting2019} and we defer a corresponding analysis and discussion to a future publication (Breton, Dyrek et al., in prep). We finally consider 34 stars in our working sample.

\begin{figure}[ht!]
    \centering
    \includegraphics[width=0.49\textwidth]{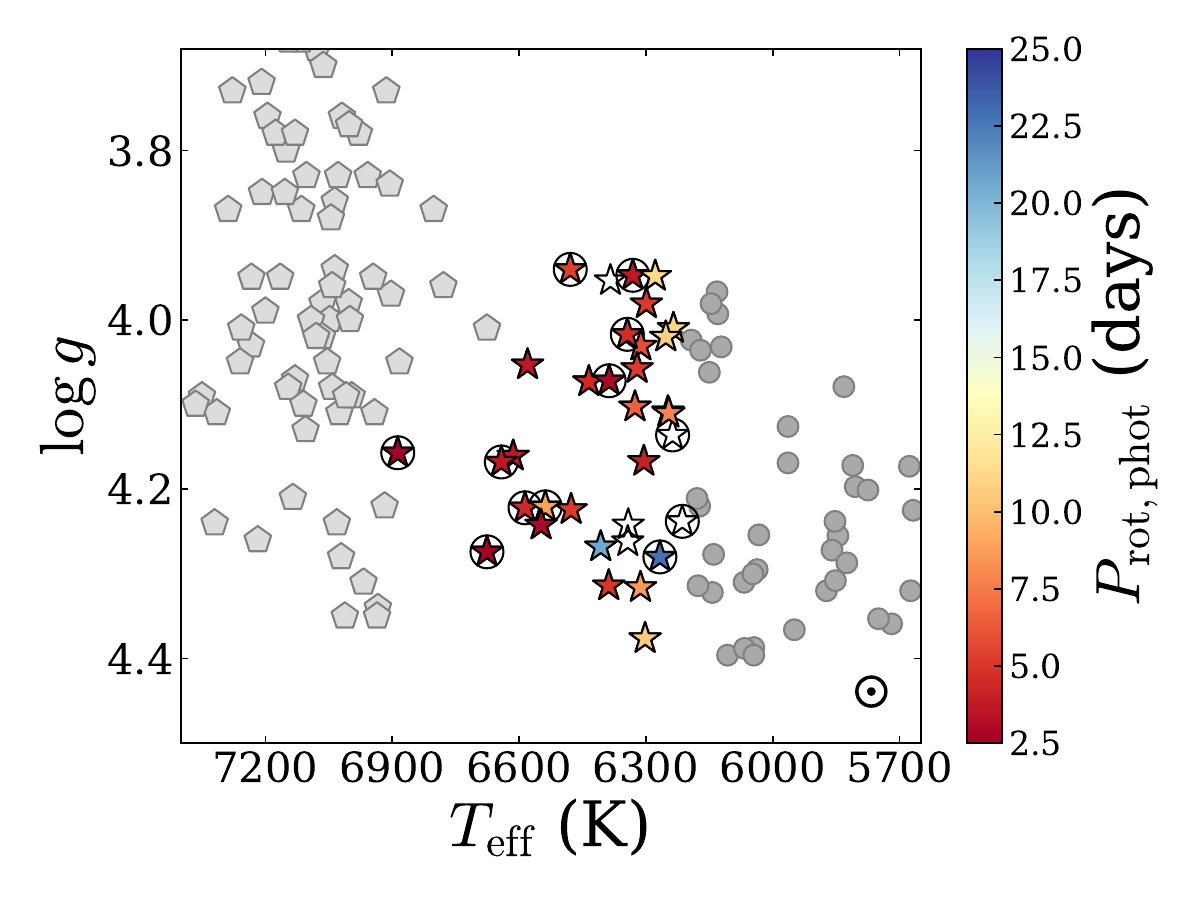}
    \caption{$T_\mathrm{eff}$ versus $\log g$ diagram for the selected sample (stars), the remaining solar-type pulsators from \citet[][grey circles,]{Lund2017} and the $\gamma$ Dor sample from \citet{VanReeth2015b} and \citet[][grey pentagons]{Gebruers2021}. The Sun is represented by the symbol $\odot$. The photometric rotation period is colour-coded for the selected sample when available; otherwise the target symbol remains unfilled. The stars for which we obtain a detection with the ensemble analysis described in Sect.~\ref{sec:ensemble_analysis} are encircled in black.
    }
    \label{fig:teff_logg_diagram}
\end{figure}

\begin{figure}[ht!]
    \centering
    \includegraphics[width=0.49\textwidth]{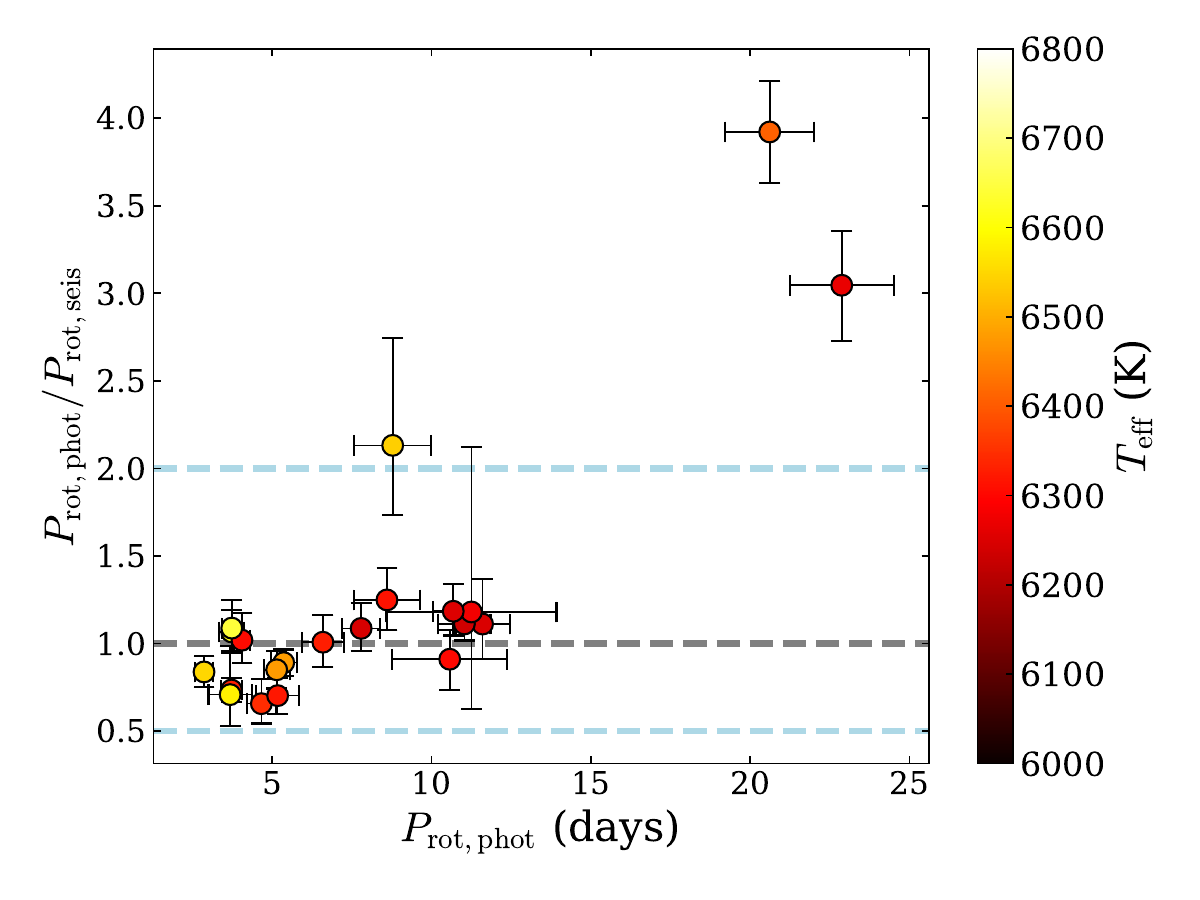}
    \caption{Comparison between $P_\mathrm{rot,phot}$ measured by \citet{Santos2021} and $P_\mathrm{rot,seis}$ measured by \citet{Hall2021} for the selected sample, with corresponding error bars. The $T_\mathrm{eff}$ values are colour coded.}
    \label{fig:prot_phot_seis}
\end{figure}

Figure~\ref{fig:teff_logg_diagram} show the $T_\mathrm{eff}$ versus $\log g$ diagram for the selected sample. When available, we also colour code the photometric average surface rotation period of the star, $P_\mathrm{rot,phot}$, as measured by \citet{Santos2021}.
In order to illustrate the location of the selected sample, we show the Sun, the stars from \citet{Lund2017} that were not included in our sample, and the $\gamma$ Dor from \citet{VanReeth2015b} and \citet{Gebruers2021}, which have been spectrocospically characterised. It is remarkable that, when excluding the case of KIC~9226926 discussed above, there is no overlap on the diagram between the solar-type pulsators and the $\gamma$ Dor stars. In particular, a thin depleted band without pulsators is visible between the two populations. Given the uncertainties on the location of the red edge of the $\gamma$~Dor instability strip \citep{Guzik2000,Dupret2004,Dupret2005,Xiong2016}, this separation raises questions about the existence of $\gamma$~Dor--solar-type hybrid pulsators in the transition region between the two populations. Before presenting our analysis strategy  in more detail in Sect.~\ref{sec:ensemble_analysis}, we underline here that none of the stars of our sample exhibit $\gamma$-Dor-like large-amplitude gravito-inertial modes. However, with almost no short-cadence \textit{Kepler} data available for the $\gamma$~Dor sample, it is unclear whether or not stochastic excitation from the thin convective layer of these stars is still able to excite detectable p modes in the coolest of them \citep[e.g.][]{Belkacem2010}.

Figure~\ref{fig:prot_phot_seis} shows a comparison of $P_\mathrm{rot,phot}$ with the seismic rotation periods $P_\mathrm{rot,seis}$ measured by \citet{Hall2021} .
As expected, the hottest and most massive stars of the sample tend to be the fastest rotators \citep[see e.g.][]{vanSaders2013,Matt2015}.
For these stars, the corresponding fundamental rotation frequency is of the order of a few $\mu$Hz, which means that their PSD is likely to exhibit a power contribution from the rotation harmonics at frequencies up to a few tens of $\mu$Hz. 
We note discrepancies between $P_\mathrm{rot,phot}$ and $P_\mathrm{rot,seis}$ for several targets. 
 For the fastest rotators, with $P_\mathrm{rot,phot}$ below $\sim$5-6~days, several photometric measurements seem below the seismic values. We expect that stars in this range have low Rossby numbers and therefore exhibit solar-type latitudinal differential rotation \citep[e.g.][]{Brun2017a,Noraz2022}. In principle, when considering only low-degree modes, the seismically inferred rotation period is sensitive to the stellar rotation at low latitudes (relative to the stellar equator), and is able to probe different layers of the star \citep[see e.g.][]{Benomar2015}.
The photometrically inferred rotation period is influenced by the inclination of the rotation axis of the star with respect to the observer and by the latitude of active regions \citep[e.g.][]{Santos2017}. This alone could create significant discrepancies between photometric and seismic techniques, as latitudinal differential rotation can reach ratios of up to 140\% between the equatorial rotation frequency and the rotation frequency at 45 degrees  latitude \citep{Benomar2018}.
We also underline that the correlation between inclination angle and rotational splitting makes it particularly difficult to obtain accurate measurements of these two parameters for stars with wide modes, such as late F-type stars \citep[e.g.][]{Ballot2006,Ballot2008,Benomar2009}. This means that the uncertainties on seismic parameters could be underestimated. 
Some uncertainties on $P_\mathrm{rot,phot}$ might also be underestimated.
Comparing the bias between the photometric and seismic measurements would finally require a careful examination of the 2D kernels of the modes used by \citet{Hall2021} to measure the average rotation period, because it is important to estimate the latitudinal and radial extent over which this average is sensitive.
Explaining the exact origin of the discrepancy between photometric and seismic rotation-period measurements is therefore out of the scope of this paper.
We finally simply underline that the specific case of KIC~8866102 might be explained by the system binarity, with the 20.6 day photometric modulation being, in reality, the rotational modulation of the K-type companion.

This work provides new stellar models for four targets of the sample, KIC~3733735, KIC~6679371, KIC~7103006, and KIC~9206432. 
The modelling was performed using the IACgrid generated with the Module for Experiments in Stellar Astrophysics \citep[MESA,][]{Paxton2011,Paxton2013,Paxton2015,Paxton2018,Paxton2019}.
The modelling procedure is extensively described in \citet{Gonzalez-Cuesta2023}; we outline its main details in Appendix~\ref{appendix:modelling}. An important uncertainty for the modelling of stars with $M_\star \geq 1.2$~$\rm M_\odot$ is related to the choices concerning convective core overshooting and processes driving chemical mixing close to the core \citep[e.g.][]{Silva-Aguirre2011,Claret2016,Pratt2017,Deal2020,Johnston2021,Baraffe2023,Varghese2023}. For the sake of simplicity concerning this aspect, our modelling is restricted in this regard to the standard MESA input physics, which does not include overshooting. In the following sections, we discuss  how this might affect the comparison between models and observations.
$M_\star$ and $R_\star$ corresponding to our modelling are used in Table~\ref{tab:fstar_selection}. For these four stars as well as the ones for which the ASTFIT modelling is available, we include in Table~\ref{tab:fstar_selection} the corresponding radius at the base of the convective envelope, $R_\mathrm{CZ}$. We note that, except for KIC~9139151, all the stars in our sample have $R_\mathrm{CZ} > 0.8$~$R_\star$, and six of them have $R_\mathrm{CZ} > 0.9$~$R_\star$ (KIC~2837475, KIC~3733735, KIC~6679371, KIC~9206432, KIC~11253226, and KIC~12317678).  
When available, we also provide the value of the photometric magnetic activity index $S_\mathrm{ph}$ \citep{Garcia2010,Mathur2014b} computed by \citet{Santos2021}. This index is expected to be low (at most a few hundreds of parts per million (ppm)) for stars where p~modes are observed \citep[see][]{Mathur2019}. We reiterate the fact that, depending on the activity cycle epoch, the solar $S_\mathrm{ph}$ varies between $\sim$90 and $\sim$260~ppm \citep[see][]{Mathur2014}. Therefore, even if the $S_\mathrm{ph}$ is only a lower-limit indicator for photospheric activity (it can be biased by stellar inclination and spot latitude), the $S_\mathrm{ph}$ values of the stars in our sample correspond to a low to moderate level of solar activity. This is interesting, as high $S_\mathrm{ph}$ values would be associated to a high level of power at low frequency in the PSD, making the detection of g~modes more difficult.
Finally, Table~\ref{tab:fstar_selection} also includes an estimate of the metallicity $\rm [Fe/H]$, as this parameter is correlated with the thickness of the convective envelope: for a given mass and radius, a star with low metallicity will have a thinner convective envelope than a star with high metallicity \citep[e.g.][]{vanSaders2013,Amard2019} due to the increase in opacity with metallicity. Table~\ref{tab:param_sismo} provides the global seismic parameters $\Delta\nu$ and the frequency at maximum amplitude of the p modes  for each star, as measured by \citet{Huber2013}, \citet{Mathur2014}, and \citet{Lund2017}.

\begin{table*}[ht!]
\centering
\caption{Parameters of the selected targets}
\begin{adjustbox}{width=\textwidth}
\begin{tabular}{lrrrrrrrrr}
\hline
\hline
KIC & $T_\mathrm{eff}$ & $\log g$ & $M_\star$  & $R_\star$ & $R_\mathrm{CZ}$ & $P_\mathrm{rot, phot}$ & $P_\mathrm{rot, seis}$ & $S_\mathrm{ph}$ & $\rm [Fe/H]$ \\
{}  &   (K)   & (dex)   & ($\rm M_\odot$) & ($\rm R_\odot$)  &  ($R_\star$) & (days)  &  (days)  &  (ppm) & (dex) \\
\hline
1430163  &        $6586 \pm 85$ &  $4.222 \pm 0.015$ &           $1.29 \pm 0.08$ &           $1.46 \pm 0.06$ &                           - &                 $4.4 \pm 0.8$ &                             - &          $208 \pm 14$ &   $-0.02 \pm 0.15$ \\
1435467  &        $6326 \pm 77$ &  $4.103 \pm 0.005$ &           $1.34 \pm 0.05$ &           $1.70 \pm 0.02$ &             $0.84 \pm 0.02$ &                 $6.6 \pm 0.7$ &           $6.5^{+0.8}_{-0.6}$ &          $196 \pm 10$ &    $0.01 \pm 0.10$ \\
2837475  &        $6614 \pm 77$ &  $4.161 \pm 0.005$ &           $1.39 \pm 0.05$ &           $1.62 \pm 0.02$ &             $0.92 \pm 0.01$ &                 $3.7 \pm 0.4$ &           $3.5^{+0.2}_{-0.2}$ &            $71 \pm 5$ &    $0.01 \pm 0.10$ \\
3425851  &        $6342 \pm 86$ &  $4.242 \pm 0.033$ &           $1.18 \pm 0.07$ &           $1.36 \pm 0.07$ &                           - &                             - &           $8.1^{+8.6}_{-2.7}$ &                     - &   $-0.06 \pm 0.10$ \\
3456181  &        $6384 \pm 77$ &  $3.954 \pm 0.002$ &           $1.56 \pm 0.03$ &           $2.18 \pm 0.02$ &             $0.88 \pm 0.01$ &                             - &          $10.7^{+2.0}_{-2.8}$ &                     - &   $-0.15 \pm 0.10$ \\
3733735  &        $6676 \pm 80$ &  $4.274 \pm 0.015$ &           $1.26 \pm 0.06$ &           $1.38 \pm 0.03$ &             $0.94 \pm 0.01$ &                 $2.6 \pm 0.2$ &                             - &          $239 \pm 20$ &   $-0.02 \pm 0.15$ \\
4349452  &        $6267 \pm 81$ &  $4.280 \pm 0.033$ &           $1.10 \pm 0.09$ &           $1.26 \pm 0.07$ &                           - &                $22.9 \pm 1.6$ &           $7.5^{+0.5}_{-0.6}$ &          $360 \pm 10$ &   $-0.06 \pm 0.15$ \\
4638884  &        $6312 \pm 82$ &  $4.031 \pm 0.012$ &           $1.38 \pm 0.13$ &           $1.88 \pm 0.10$ &                           - &                 $6.0 \pm 0.7$ &                             - &           $114 \pm 7$ &    $0.10 \pm 0.15$ \\
5371516  &        $6299 \pm 88$ &  $3.981 \pm 0.011$ &           $1.31 \pm 0.11$ &           $1.94 \pm 0.09$ &                           - &                 $5.2 \pm 0.5$ &                             - &          $318 \pm 19$ &    $0.04 \pm 0.15$ \\
6225718  &        $6313 \pm 77$ &  $4.316 \pm 0.003$ &           $1.15 \pm 0.03$ &           $1.23 \pm 0.01$ &             $0.81 \pm 0.01$ &                 $8.6 \pm 1.0$ &           $6.9^{+0.6}_{-0.5}$ &            $24 \pm 1$ &   $-0.07 \pm 0.10$ \\
6508366  &        $6331 \pm 77$ &  $3.947 \pm 0.002$ &           $1.58 \pm 0.03$ &           $2.21 \pm 0.02$ &             $0.87 \pm 0.01$ &                 $3.7 \pm 0.3$ &           $5.1^{+0.1}_{-0.1}$ &          $231 \pm 16$ &   $-0.05 \pm 0.10$ \\
6679371  &        $6479 \pm 77$ &  $3.941 \pm 0.003$ &           $1.50 \pm 0.01$ &           $2.18 \pm 0.01$ &             $0.91 \pm 0.01$ &                 $5.4 \pm 0.4$ &           $6.0^{+0.2}_{-0.2}$ &           $108 \pm 7$ &    $0.01 \pm 0.10$ \\
7103006  &        $6344 \pm 77$ &  $4.017 \pm 0.004$ &           $1.47 \pm 0.02$ &           $1.96 \pm 0.01$ &             $0.87 \pm 0.01$ &                 $4.7 \pm 0.5$ &           $7.1^{+1.3}_{-1.0}$ &          $377 \pm 23$ &    $0.02 \pm 0.10$ \\
7206837  &        $6305 \pm 77$ &  $4.167 \pm 0.005$ &           $1.33 \pm 0.04$ &           $1.57 \pm 0.02$ &             $0.84 \pm 0.02$ &                 $4.1 \pm 0.3$ &           $4.0^{+0.6}_{-0.4}$ &          $234 \pm 16$ &    $0.10 \pm 0.10$ \\
7670943  &       $6477 \pm 116$ &  $4.224 \pm 0.033$ &           $1.30 \pm 0.07$ &           $1.46 \pm 0.09$ &                           - &                 $5.1 \pm 0.4$ &           $6.0^{+0.6}_{-0.6}$ &            $13 \pm 3$ &    $0.10 \pm 0.15$ \\
7771282  &        $6248 \pm 77$ &  $4.109 \pm 0.007$ &           $1.25 \pm 0.06$ &           $1.63 \pm 0.03$ &             $0.83 \pm 0.02$ &                $11.6 \pm 0.9$ &          $10.4^{+2.3}_{-1.7}$ &            $81 \pm 4$ &   $-0.02 \pm 0.10$ \\
7940546  &        $6235 \pm 77$ &  $4.010 \pm 0.002$ &           $1.48 \pm 0.03$ &           $1.99 \pm 0.01$ &             $0.84 \pm 0.01$ &                $11.0 \pm 0.8$ &           $9.9^{+0.3}_{-0.4}$ &            $52 \pm 2$ &   $-0.20 \pm 0.10$ \\
8179536  &        $6343 \pm 77$ &  $4.262 \pm 0.006$ &           $1.20 \pm 0.04$ &           $1.34 \pm 0.02$ &             $0.84 \pm 0.02$ &                             - &           $6.5^{+1.2}_{-0.8}$ &                     - &   $-0.03 \pm 0.10$ \\
8292840  &       $6214 \pm 113$ &  $4.238 \pm 0.033$ &           $1.06 \pm 0.09$ &           $1.30 \pm 0.07$ &                           - &                             - &           $7.7^{+0.5}_{-0.5}$ &                     - &   $-0.16 \pm 0.10$ \\
8694723  &        $6246 \pm 77$ &  $4.110 \pm 0.002$ &           $1.11 \pm 0.02$ &           $1.53 \pm 0.01$ &             $0.82 \pm 0.01$ &                 $7.8 \pm 0.6$ &           $7.2^{+0.8}_{-0.6}$ &            $29 \pm 2$ &   $-0.42 \pm 0.10$ \\
8866102  &        $6407 \pm 83$ &  $4.268 \pm 0.030$ &           $1.21 \pm 0.09$ &           $1.34 \pm 0.07$ &                           - &                $20.6 \pm 1.4$ &           $5.3^{+0.2}_{-0.2}$ &           $274 \pm 8$ &    $0.00 \pm 0.15$ \\
9139151  &        $6302 \pm 77$ &  $4.376 \pm 0.004$ &           $1.15 \pm 0.03$ &           $1.15 \pm 0.01$ &             $0.78 \pm 0.01$ &                $10.6 \pm 1.8$ &          $11.6^{+0.8}_{-1.1}$ &           $170 \pm 7$ &    $0.10 \pm 0.10$ \\
9206432  &        $6538 \pm 77$ &  $4.221 \pm 0.005$ &           $1.32 \pm 0.04$ &           $1.47 \pm 0.02$ &             $0.92 \pm 0.01$ &                 $8.8 \pm 1.2$ &           $4.1^{+1.0}_{-0.5}$ &            $53 \pm 3$ &    $0.16 \pm 0.10$ \\
9226926  &        $6887 \pm 89$ &  $4.157 \pm 0.065$ &           $1.34 \pm 0.07$ &           $1.60 \pm 0.14$ &                           - &               $2.17 \pm 0.14$ &                             - &      $103.7 \pm 5.4$ &   $-0.22 \pm 0.10$ \\
9353712  &        $6278 \pm 77$ &  $3.948 \pm 0.003$ &           $1.56 \pm 0.04$ &           $2.19 \pm 0.03$ &             $0.87 \pm 0.02$ &                $11.2 \pm 2.7$ &           $9.5^{+7.3}_{-3.9}$ &            $47 \pm 3$ &   $-0.05 \pm 0.10$ \\
9414417  &        $6253 \pm 75$ &  $4.020 \pm 0.002$ &           $1.45 \pm 0.03$ &           $1.95 \pm 0.02$ &             $0.84 \pm 0.01$ &                $10.7 \pm 0.6$ &           $9.0^{+1.1}_{-0.9}$ &            $89 \pm 4$ &   $-0.13 \pm 0.10$ \\
9812850  &        $6321 \pm 77$ &  $4.057 \pm 0.003$ &           $1.37 \pm 0.04$ &           $1.81 \pm 0.02$ &             $0.85 \pm 0.01$ &                 $5.2 \pm 0.7$ &           $7.4^{+0.4}_{-0.5}$ &          $196 \pm 12$ &   $-0.07 \pm 0.10$ \\
10016239 &        $6388 \pm 89$ &  $4.314 \pm 0.010$ &           $1.18 \pm 0.08$ &           $1.25 \pm 0.04$ &                           - &                 $4.9 \pm 0.5$ &                             - &            $65 \pm 5$ &   $-0.02 \pm 0.15$ \\
10355856 &        $6435 \pm 83$ &  $4.073 \pm 0.010$ &           $1.21 \pm 0.12$ &           $1.67 \pm 0.09$ &                           - &                 $4.5 \pm 0.3$ &                             - &          $300 \pm 20$ &   $-0.10 \pm 0.15$ \\
11070918 &       $6387 \pm 192$ &  $4.072 \pm 0.023$ &           $1.16 \pm 0.20$ &           $1.64 \pm 0.15$ &                           - &                 $2.9 \pm 0.2$ &                             - &          $136 \pm 12$ &   $-0.20 \pm 0.25$ \\
11081729 &        $6548 \pm 83$ &  $4.242 \pm 0.006$ &           $1.32 \pm 0.05$ &           $1.44 \pm 0.02$ &             $0.86 \pm 0.02$ &                 $2.9 \pm 0.3$ &           $3.4^{+0.1}_{-0.1}$ &          $271 \pm 22$ &    $0.11 \pm 0.10$ \\
11253226 &        $6642 \pm 77$ &  $4.168 \pm 0.005$ &           $1.37 \pm 0.05$ &           $1.59 \pm 0.02$ &             $0.92 \pm 0.01$ &                 $3.7 \pm 0.3$ &           $3.4^{+0.4}_{-0.3}$ &            $67 \pm 5$ &   $-0.08 \pm 0.10$ \\
11807274 &        $6237 \pm 74$ &  $4.136 \pm 0.033$ &           $1.19 \pm 0.08$ &           $1.54 \pm 0.08$ &                           - &                             - &           $7.9^{+0.5}_{-0.5}$ &                     - &    $0.00 \pm 0.10$ \\
12317678 &        $6580 \pm 77$ &  $4.053 \pm 0.003$ &           $1.37 \pm 0.03$ &           $1.82 \pm 0.02$ &             $0.91 \pm 0.01$ &                 $3.7 \pm 0.7$ &           $5.2^{+1.8}_{-0.9}$ &            $51 \pm 4$ &   $-0.28 \pm 0.10$ \\
\hline
\end{tabular}
\end{adjustbox}
\tablefoot{$T_\mathrm{eff}$, $\log g$, $M_\star$, and $R_\star$ values are from the DR25 catalogue \citep{Mathur2017}. When possible, $\log g$, $M_\star$, and $R_\star$ values are replaced by the values from the IACgrid \citep{Perez-Hernandez2016,Perez-Hernandez2019,Gonzalez-Cuesta2023} modelling (KIC~3733735, KIC~6679371, KIC~7103006, and KIC~9206432) or the ASTFIT modelling from \citet{SilvaAguirre2017}. In this case, $T_\mathrm{eff}$ and $\rm [Fe/H]$ values are replaced by those from \citet{Lund2017}. For these stars, the radius at the base of the convective zone $R_\mathrm{CZ}$  is provided. $P_\mathrm{rot, phot}$ and $S_\mathrm{ph}$ are taken from \citet{Santos2019,Santos2021}, with the exception of KIC~9226926, for which we use the value from \citet{Mathur2014}. $P_\mathrm{rot, seis}$ are taken from \citet{Hall2021}. 
}
\label{tab:fstar_selection}
\end{table*}

\begin{table}[ht!]
\centering
\caption{Frequency of maximum amplitude $\nu_\mathrm{max}$ and large frequency separation $\Delta\nu$ for the stars in the sample.}
\begin{tabular}{lllc}
\hline
\hline
KIC & $\nu_\mathrm{max}$ ($\mu$Hz) & $\Delta\nu$ ($\mu$Hz) & Origin \\
\hline
1430163  &               $1805.07\pm29.66$ &        $85.66\pm1.80$   & \citetalias{Mathur2014} \\
1435467  &               $1406.7\pm8.4$ &           $70.369\pm0.034$ & \citetalias{Lund2017} \\
2837475  &               $1557.6\pm9.2$ &        $75.729\pm0.042$ & \citetalias{Lund2017} \\
3425851  &               $2038\pm60$ &        $92.6\pm1.5$ & \citetalias{Huber2013} \\
3456181  &                $970.0\pm8.3$ &        $52.264\pm0.041$ & \citetalias{Lund2017} \\
3733735  &               $2132.61\pm84.44$ &        $92.37\pm1.72$ & \citetalias{Mathur2014} \\
4349452  &               $2106\pm50$ &        $98.27\pm0.57$ & \citetalias{Huber2013} \\
4638884  &               $1192.28\pm75.82$ &        $60.83\pm1.15$ & \citetalias{Mathur2014} \\
5371516  &               $1018.62\pm59.84$ &        $55.46\pm1.06$ & \citetalias{Mathur2014} \\
6225718  &               $2364.2\pm4.9$ &       $105.695\pm0.018$ & \citetalias{Lund2017} \\
6508366  &                $958.3\pm4.6$ &        $51.553\pm0.047$ & \citetalias{Lund2017} \\
6679371  &                $941.8\pm5.1$ &        $50.601\pm0.029$ & \citetalias{Lund2017} \\
7103006  &               $1167.9\pm7.2$ &        $59.658\pm0.030$ & \citetalias{Lund2017} \\
7206837  &               $1652.5\pm11.7$ &        $79.131\pm0.039$ & \citetalias{Lund2017} \\
7670943  &               $1895\pm73$ &        $88.6\pm1.3$ & \citetalias{Huber2013} \\
7771282  &               $1465.1\pm27.0$ &       $72.463\pm0.079$ & \citetalias{Lund2017} \\
7940546  &               $1116.6\pm3.6$ &         $58.762\pm0.029$ & \citetalias{Lund2017} \\
8179536  &               $2074.9\pm13.8$ &        $95.090\pm0.058$ & \citetalias{Lund2017} \\
8292840  &               $1983\pm37$ &        $92.85\pm0.35$ & \citetalias{Huber2013} \\
8694723  &               $1470.5\pm4.1$ &        $75.112\pm0.021$ & \citetalias{Lund2017} \\
8866102  &               $2014\pm32$ &        $94.50\pm0.60$ & \citetalias{Huber2013} \\
9139151  &               $2690.4\pm14.5$ &       $117.294\pm0.032$ & \citetalias{Lund2017} \\
9206432  &               $1866.4\pm14.9$ &        $84.926\pm0.051$ & \citetalias{Lund2017} \\
9226926  &               $1411.43\pm0.99$ &        $73.77\pm1.26$ & \citetalias{Mathur2014} \\
9353712  &                $934.3\pm11.1$ &        $51.467\pm0.104$ & \citetalias{Lund2017} \\
9414417  &               $1155\pm32$ &        $60.05\pm0.27$ & \citetalias{Huber2013} \\
9812850  &               $1255.2\pm9.1$ &        $64.746\pm0.068$ & \citetalias{Lund2017} \\
10016239 &               $2087.74\pm22.83$ &       $101.60\pm2.59$ & \citetalias{Mathur2014} \\
10355856 &               $1298.44\pm64.19$ &        $67.47\pm1.43$ & \citetalias{Mathur2014} \\
11070918 &               $1093.64\pm20.03$ &        $66.89\pm1.61$ & \citetalias{Mathur2014} \\
11081729 &               $1968.3\pm12.6$ &        $90.116\pm0.048$ & \citetalias{Lund2017} \\
11253226 &               $1590.6\pm10.6$ &        $76.858\pm0.030$ & \citetalias{Lund2017} \\
11807274 &               $1496\pm56$ &        $75.71\pm0.31$ & \citetalias{Huber2013} \\
12317678 &               $1212.4\pm5.5$ &        $63.464\pm0.025$ & \citetalias{Lund2017} \\
\hline
\end{tabular}
\tablefoot{In the following order of priority
for stars present in several lists, the seismic parameters are taken from \citet[][L17]{Lund2017}, \citet[][M14]{Mathur2014}, and \citet[][H13]{Huber2013}. For stars for which lower and upper uncertainties are provided, we retain only the largest value.}
\label{tab:param_sismo}
\end{table}

\section{Ensemble analysis \label{sec:ensemble_analysis}}

\subsection{Estimating the background power level \label{sec:background_estimation}}

In the frequency range where g modes are expected for the considered stars, the PSD is dominated by stellar magnetic activity, granulation, and surface rotation peaks due to dark spots and faculae \citep[see e.g.][and references therein]{Garcia2019}. Before looking for significant peaks in the spectrum, it is therefore necessary to remove the background contribution \citep[e.g.][]{Mathur2011,Kallinger2014} in the PSD while limiting the bias introduced by the presence of rotational harmonics.

Extreme low-frequency regions of the PSD can be dominated by long-term activity variations, such as magnetic cycles, and/or still be polluted by instrumental modulations even after applying the 55 day high-pass filtering. Low-frequency regions exhibit a significant contribution from rotational-modulation harmonics. Therefore, we consider only frequency peaks above 20~$\mu$Hz  for our fit. 
The following background model is used
\begin{equation}
\label{eq:B_p}
    B (\nu) = a \nu^{-b} + h \; ,
\end{equation}
where $a$ and $b$ are the parameters of the power law modelling the low-frequency trend of the PSD. The parameter $h$ accounts for the noise component close to the Nyquist frequency, which for long-cadence data is dominated by stellar convective granulation.
We emphasise that this low-frequency background model corresponds to the modelling choice of \citet{Mathur2010}, and that in order to model the lowest-frequency region of the PSD, \citet{Aigrain2004} or \citet{Kallinger2014} preferred the use of a Harvey profile \citep{Harvey1985} with a low-frequency cutoff rather than a power law. 
It indeed remains unclear as to which is the optimal way to model this low-frequency area corresponding to the magnetic activity signature in the Fourier spectrum.
Nevertheless, as we use a 20~$\mu$Hz low-frequency cutoff, which excludes a significant part of the low-frequency power slope, the power-law parameterisation of our model provides more flexibility for the fit.

\begin{figure}[ht!]
    \centering
    \includegraphics[width=0.48\textwidth]{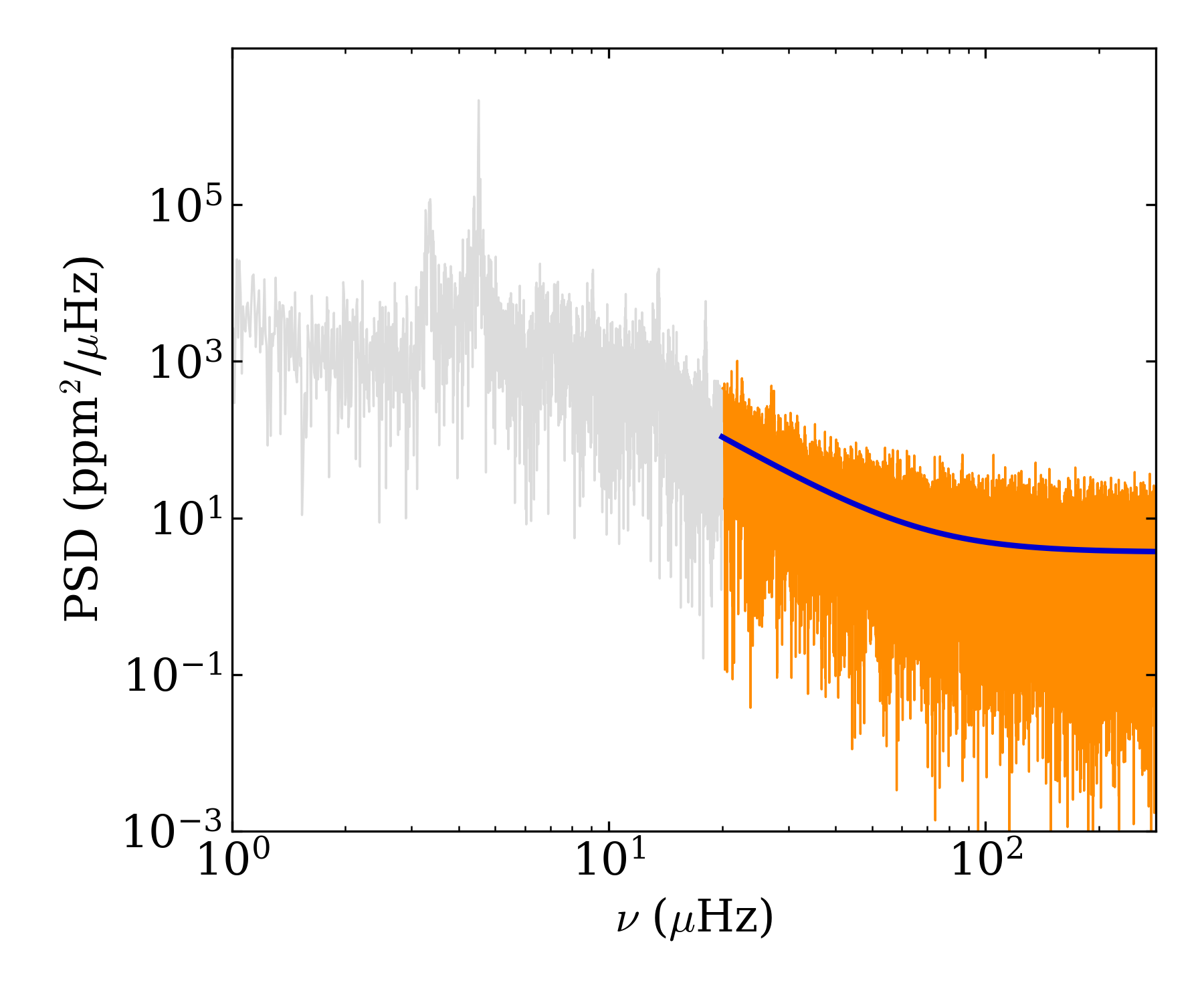}
    \caption{Background model computed for KIC~3733735. Only the frequency bins in dark orange are considered for the fit. 
    The low-frequency area of the PSD (in grey) is not considered when computing the likelihood.
    In blue, we show the background profile $B$ we obtain from the posterior probability MCMC sampling.    
    }
    \label{fig:fit_background_parametric}
\end{figure}

The Markov Chain Monte Carlo (MCMC) exploration of the parameter distribution is performed with the \texttt{apollinaire} module\footnote{The module documentation is available at \url{apollinaire.readthedocs.io/en/latest}.} \citep[see][for a complete description of the module]{Breton2022apollinaire}, assuming that the probability distribution of the likelihood $\mathcal{L}$ follows a $\chi^2$ with two degrees of freedom
\begin{equation}
    \mathcal{L} (\mathrm{PSD}, \theta) = \prod_k \frac{1}{B(\nu_k, \theta)} \exp \left[ - \frac{\mathrm{PSD}(\nu_k)}{B(\nu_k, \theta)} \right] \; . 
\end{equation}
The posterior probability distribution we want to sample is therefore 
\begin{equation}
    p (\theta, \mathrm{PSD})= \mathcal{L} (\mathrm{PSD}, \theta) p (\theta) \; ,
\end{equation}
where $p (\theta)$ is the prior distribution assumed for the set of parameters $\theta$. 
Here, we choose to sample $\theta$ = ($\ln a$, $\ln b$, $\ln h$). The choice to sample the logarithm of the parameter allows us to use uniform distributions as non-informative priors \citep{Benomar2009}, ensuring with an a posteriori inspection of the sampled distributions that the chosen bounds are wide enough to avoid introducing a bias in the inferred posterior probability distribution. 
We sample the posterior probability with 32 walkers and 10000 steps ---from which we discard the first 5000 steps in order to ensure that the sampled posterior is not biased by the presence of a tail resulting from the exploration walk from the initial position of the walker to the high-probability regions. 
We illustrate our method in Fig.~\ref{fig:fit_background_parametric} with the case of KIC~3733735, which exhibits a rotational modulation of particularly large amplitude and an important number of harmonics. It is possible to see that, by setting the 20~$\mu$Hz low-frequency cutoff, no apparent bias related to the power excess of the rotational modulation affects the background profile we obtain from our MCMC sampling. 

In order to perform the statistical analysis described in Sect.~\ref{sec:stat_analysis}, we now use the background model to obtain the signal-to-noise-ratio(S/N) spectrum, $\mathrm{P}_{\rm S/N}$
\begin{equation}
    \mathrm{P}_{\rm S/N} \, (\nu) = \frac{\mathrm{PSD} \, (\nu)}{B(\nu)} \; .
\end{equation}

\subsection{Statistical analysis \label{sec:stat_analysis}}

Given an oscillation mode of frequency $\nu$ and unknown lifetime, it is not trivial to define its S/N detection threshold. This matter has been discussed for example by \citet{Appourchaux2000}, who, under the assumption of a distribution of $\chi^2$   with two degrees of freedom, suggested the following formula for the detection threshold of a signal of power $s_\mathrm{det}$ in continuous disc-integrated observations:
\begin{equation}
\label{eq:appourchaux2000}
    \frac{s_\mathrm{det}}{B (\nu)} \approx \ln t_\mathrm{obs} + \ln \delta\nu - \ln p_\mathrm{det}  \; ,
\end{equation}
where $t_\mathrm{obs}$ is the observing time, $\delta\nu$ is the considered frequency interval, and $p_\mathrm{det} \ll 1$ is the probability that at least one peak in the $\delta \nu$ window is above $s_\mathrm{det}$ in the null hypothesis, H0. For $p_\mathrm{det} = 0.1$, the value adopted by \citet{Appourchaux2000} and \citet{Belkacem2009}, a four-year time series, and a $\delta \nu$ window of $\sim 263$~$\mu$Hz (i.e. the width of the band that we are exploring between the low-frequency cutoff and the Nyquist frequency), this formula yields a detection threshold of $s_\mathrm{det} \approx 12.7 \, B(\nu)$. In other words, this means that, if the power peaks in the $\delta\nu$ window are exactly distributed following a $\chi^2$ distribution, there is just a 10\% probability that a peak appears above the threshold as a pure noise feature. 

We emphasise that what is interpreted here as a probable signal is a deviation from the assumed $\chi^2$ distribution, which may have different possible origins, including instrumental artefacts. 
A detection of significant peaks needs to be strengthened by characterisation of the intrinsic properties of the phenomenon that we are trying to detect. In the case of g modes, the $\Delta P_\ell$ asymptotic regular period spacing derived by \citet{Tassoul1980} appears to be a reasonable choice for this characterisation.
However, when looking for individual significant peaks, this requires the detection of at least several significant peaks in order to reconstruct the pattern and assess its regularity, which is a complex task in the case of low-S/N oscillation modes.  
An alternative solution is to look for the global contribution to the pattern of regularly spaced non-individually significant peaks, as chosen by \citet{Garcia2007} in order to highlight evidence of the solar g-mode signature. However, transposed to a stellar case, this method has a large degeneracy that needs to be examined carefully (stellar inclination angle, influence of rotational splittings, degree-dependence of the pattern), and it is therefore beyond the scope of this paper to apply this method to our sample.

High-order (low-frequency) g mode period spacing may also deviate from the Tassoul asymptotic relation under the effect of rotation \citep[e.g.][]{Bouabid2013} or because of the steep profile of the Brunt-Väisälä frequency, $N$ at the boundary between the convective core and the radiative interior \citep[e.g.][]{Berthomieu1988}. We reiterate that $N$ is given by
\begin{equation}
    N^2 = g \left ( \frac{1}{\Gamma_1 P} \diffp{P}{r} - \frac{1}{\rho} \diffp{\rho}{r} \right) \; ,
\end{equation}
where $g$, $P,$ and $\rho$ are the gravity acceleration, pressure, and density of the medium, respectively, while $\Gamma_1 = (\partial \ln P / \partial \ln \rho)_\mathrm{ad}$ is the adiabatic exponent. $N$ can be written as 
\begin{equation}
\label{eq:sum_Nt_Nmu}
    N^2 = N_T^2 + N_\mu^2 \; , 
\end{equation}
which is the sum of an entropy-stratification component, $N_T$, and a chemical-stratification component, $N_\mu$, \citep{Aerts2019}.
In more massive stars exhibiting unstable gi modes with large amplitude, the peak significance validation is only a marginal issue and it is therefore more straightforward to reconstruct the observed pattern and its deviation to the asymptotic formula. Due to fast rotation, the main challenge in this case is the correct attribution of the azimuthal number $m$ of the observed components \citep[e.g.][]{VanReeth2015b,Christophe2018}. Finally, given the respective profiles of $N$ and the Lamb frequency, $S_\ell$, in the frequency region we are exploring in this work, we can expect modes to exhibit a mixed gravito-acoustic behaviour \citep{Shibahashi1979}. This is particularly interesting as these modes have demonstrated their ability to probe structure and chemical mixing \citep[e.g.][]{Deheuvels2010,Cunha2015,Vrard2022}, rotation \citep[e.g.][]{Deheuvels2014}, and magnetism \citep[e.g.][]{Loi2020,Bugnet2021} in the deep layers of evolved stars.

We remind the reader that $S_\ell$ is connected to the speed of sound $c_s = \sqrt{\Gamma_1 P / \rho}$ through the relation
\begin{equation}
    S_\ell^2 = \frac{\ell(\ell+1)}{r^2} c_s^2  \; .
\end{equation}

Given all these considerations, we propose to start by identifying peaks of interest, and then to interpret the pattern of peaks we observe with respect to the frequency range where they are detected. 
We can then compute the probability that $k$ peaks ---with all their S/N heights $s/B \geq z$--- are a product of noise, by considering the probability $p_0$ that, in the absence of signal, there are at least $k$ peaks above $z:$
\begin{equation}
\label{eq:proba_k_peaks}
    p_0 (k | s/B(\nu)\geq z) = 1 - \sum_{i=0}^{k-1} {n_\mathrm{bin} \choose k} \exp (-iz) \Big[1 - \exp (-z)\Big]^{n_\mathrm{bin}-i} \; ,
\end{equation}
where $n_\mathrm{bin} \sim \delta\nu t_\mathrm{obs}$ is the number of frequency bins in the considered interval. Therefore, a small $p_0$ value means that it is unlikely that the observed signal is a product of noise following a $\chi^2$ with two degrees of freedom. It should be noted that, setting $k=1$ and assuming $\exp (-z) \ll 1$, Eq.~(\ref{eq:proba_k_peaks}) yields the detection threshold provided in Eq.~(\ref{eq:appourchaux2000}).

Considering the S/N spectra computed from the $B$ profile as described in Sect.~\ref{sec:background_estimation}, from 20~$\mu$Hz to the long-cadence Nyquist frequency, 283.2~$\mu$Hz, we use Eq.~(\ref{eq:appourchaux2000}) to list the peaks located above the detection threshold for each of our targets. 
The impact of rotation on the period spacing of the mode we are looking for should be limited. Indeed, the shortest rotation period of our sample (see Table~\ref{tab:fstar_selection}) is 2.17 days for KIC~9226926, which corresponds to a frequency $\Omega/2\pi \sim 5.33$~$\mu$Hz. \citet{Aerts2019} underlined that a treatment of rotation going beyond the perturbative approach \citep[e.g. the traditional approximation of rotation; see e.g.][]{Lee1997} has to be considered when $\omega < 4 \Omega$, where $\omega = 2\pi \nu$ \citep[see also][for the limit of the perturbative approach in uniformly rotating polytropes]{Ballot2010}. In \citetalias{Breton2022simuFstars}, departures from the mode frequencies yielded by perturbative predictions were observed in 3D simulations for $\omega < 10 \Omega$. Using an equatorial model, \citet{Mathis2023Tunneling} showed that the mode behaviour started to be modified with respect to the asymptotic non-rotating approximation below this same threshold.

We assume here that the resonant cavity of the mode in the radiative interior has a rotation rate that is of the same order of magnitude as that of the convective envelope.
This is a reasonable assumption if we consider that the seismic measurements in early F-type stars provide a core-to-surface rotation ratio of close to one \citep[see in particular Fig.~6 from][and references therein]{Aerts2021}, meaning that strong angular momentum-transport mechanisms are at work in the radiative layers of these stars \citep{Ouazzani2019}. Concerning solar-type pulsators,
\citet{Benomar2015} compared the convective envelope and upper layers of the radiative interior rotation rates, showing that there was no evidence of the existence of a significant amount of differential radial rotation. It should also be noted that F-type stars with a convective envelope are expected to have a very short radiative interior--convective envelope coupling timescale compared to cooler stars \citep{Spada2020}.

We find at least one significant peak in 12 stars among the 34 of our sample. Given the $p_\mathrm{det} = 0.1$ value we set, we would expect to have about $3.4 \pm 1.8$ false detections, which means that finding such a signal in 12 stars constitutes a significant detection. We also underline that this nevertheless signifies that between one-quarter and one-half of the sample may be false positives.

The list of peaks of interest we detect is provided in Table~\ref{tab:significant_peaks}, with corresponding identifier, frequency, and periods. The stars where there is a detection are encircled in Fig.~\ref{fig:teff_logg_diagram}. We note that they are not concentrated in a specific subdivision of the $T_\mathrm{eff}$ versus $\log g$ diagram inside our working sample. 
We visually inspect the S/N spectra of the 12 stars where a significant peak is found with the aim being to identify regular patterns surrounding these peaks.
Following our visual inspection, we find four targets of interest. We discuss the case of KIC~3733735 in Sect.~\ref{sec:kic3733735}, where we argue that we see both the signature of the $\ell=1$, $n=1$ p mode and a pattern of pure g modes. 
In Sect.~\ref{sec:mixed_modes}, we discuss the cases of KIC~6679371, KIC~7103006, and KIC~9206432, where the signature we observe can be explained by invoking a coupling between p- and g-mode resonant cavities.

\begin{table}
\centering
\caption{Individual peaks detected with the $p_\mathrm{det} = 0.1$ threshold.}
\label{tab:significant_peaks}
\begin{tabular}{ccc}
\hline \hline
     KIC & $\nu$ ($\mu$Hz) & Period (min) \\
\hline
 1430163 &            68.7 &   242.8 \\
 3733735 &           105.0 &   158.7 \\
 4349452 &           181.3 &    91.9 \\
 4349452 &           269.6 &    61.8 \\
 6508366 &            79.2 &   210.5 \\
 6679371 &           191.9 &    86.8 \\
 7103006 &           160.0 &   104.1 \\
 8292840 &           239.3 &    69.6 \\
 9206432 &           258.6 &    64.5 \\
 9226926 &            26.3 &   633.0 \\
11070918 &            35.2 &   473.3 \\
11253226 &           280.9 &    59.3 \\
11807274 &           132.5 &   125.7 \\
\end{tabular}
\end{table}

\section{Decoupled cavities: The case of KIC~3733735 \label{sec:kic3733735}}

Around the 158.7~min peak detected from Eq.~(\ref{eq:appourchaux2000}), KIC~3733735 exhibits five peaks with S/N above 10.
Using Eq.~(\ref{eq:proba_k_peaks}), we compute the $p_0$ probability of observing these six peaks in a $\chi^2$ distribution, setting the height of the lowest amplitude peak of the pattern for $z$, and still considering a wide frequency window of 263.2~$\mu$Hz. We could of course restrict the frequency window around the frequency interval where we observe the pattern, but keeping the initial window on which we performed the homogeneous search of individual peaks allows us to estimate more conservative values to compute the $p_0$ values.
We find $p_0 = \num{1.2e-3}$.
It should also be noted that the 158.7~min peak is above the $p_\mathrm{det} = 0.05$ threshold.

\begin{figure}[ht!]
    \centering
    \includegraphics[width=0.48\textwidth]{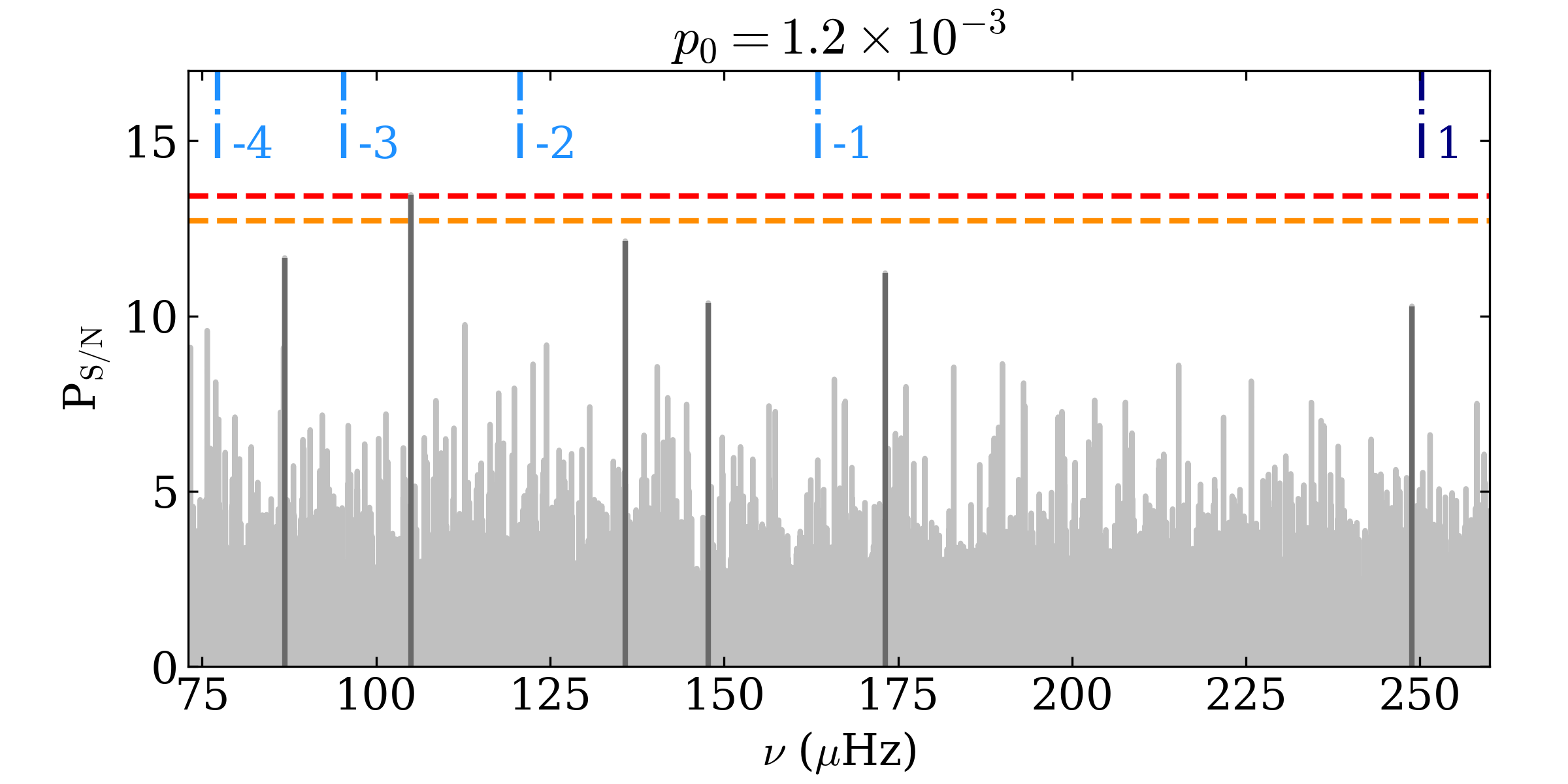}
    \includegraphics[width=0.48\textwidth]{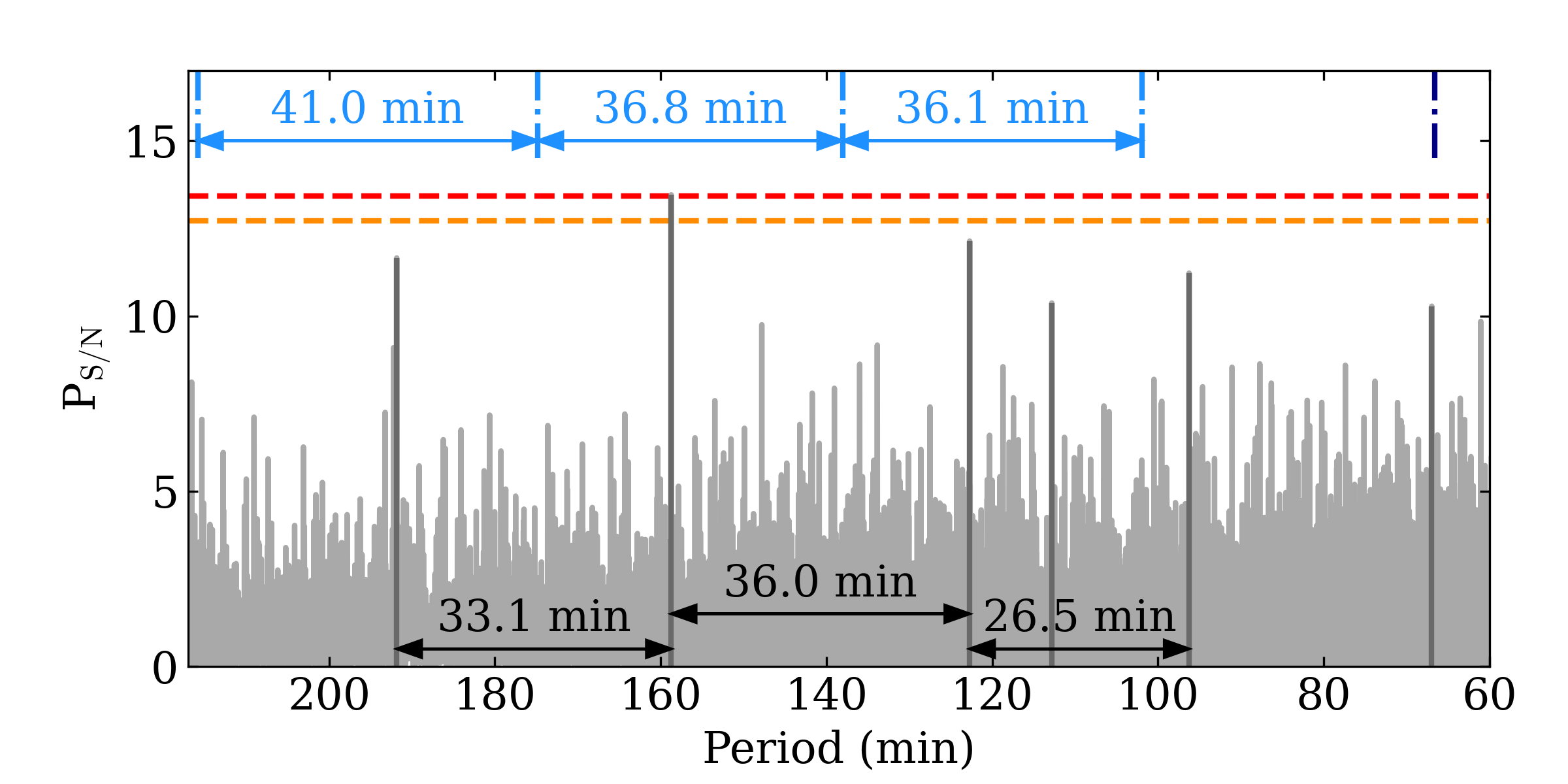}
    \caption{
    $\mathrm{P}_\mathrm{S/N}$ (\textit{grey}) for KIC~3733735, with the x-axis scaled in frequency (\textit{top}) and in period (\textit{bottom}). 
    The $p_0$ probability of the pattern is shown in the top panel.
    The dashed horizontal orange and red lines correspond to the $p_\mathrm{det} = 0.1$ and 0.05 detection thresholds computed from Eq.~(\ref{eq:appourchaux2000}), respectively.
    The peaks included when computing the pattern probability $p_0$ are shown in darker grey.
    The black arrows show the period spacing between some peaks of the pattern. The mode frequencies computed for $\ell=1$, $m=0$ modes with GYRE are shown for  (from right to left, $n=-1$, $-2$, $-3$ and $-4$) g modes (light blue) and the $n=1$ p mode (dark blue). The period spacing between consecutive g modes is shown.
    } 
    \label{fig:3733735_freq_and_period}
\end{figure}

Using our reference stellar model for KIC~3733735, we use the GYRE oscillation code \citep{Townsend2013,Townsend2018,Goldstein2020} to compute the expected oscillations frequencies for $\ell=1$ modes with low absolute order, $|n|$, following the usual convention of labelling p modes with $n > 0$ and g modes with $n < 0$.
As illustrated in Fig.~\ref{fig:3733735_freq_and_period}, the highest frequency peak, at 248.8~$\mu$Hz, closely matches the frequency predicted by GYRE for the  $\ell=1$, $n = 1$ p mode of the stellar model we compute for KIC~3733735 (250.2~$\mu$Hz). 
At lower frequency, three of the peaks are distributed following a period spacing of close to 30 min. There is one additional peak with S/N above 10 that is not included in this regular pattern. 

We note that the period spacing we observe in the signal is significantly different from what is predicted by our model. In order to discuss this point, we must remind the reader that, as already stated in Sect.~\ref{sec:sample_selection}, the reference models were computed using a grid of stellar models with standard input physics that do not account for core overshooting. The discrepancies between the observed and modelled frequencies can be explained by the fact that the asymptotic period spacing from \citet{Tassoul1980} is a function of the $\int N / r \mathrm{d}r$ integral. It is therefore extremely sensitive to the extent of the convective core and to the chemical buoyancy term, $N_\mu$ (see Eq.~(\ref{eq:sum_Nt_Nmu})), at the interface between the convective core and the radiative interior. Given the uncertainties on the overshooting and internal mixing processes  at stake, accurately modelling the $N$ profile at the convective core interface in late F-type stars represents an additional difficulty compared to lower-mass solar-type stars with a radiative core. This interface is supposed to build a strong chemical-composition gradient, as \citet{Varghese2023} found that the mixing efficiency close to the convective core decreases with stellar mass. We therefore underline that (i) the mode identification we report in the present work is achieved within the hypothesis assumed for the grid of stellar models we use for our seismic modelling, and (ii) it can be used to calibrate core overshooting and mixing in a range of masses where observational evidence is still lacking \citep[e.g.][for the case of more massive stars]{Neiner2012,Mombarg2020,Pedersen2021}. In what follows, we present the properties of the identified modes  in more detail.

\begin{figure}[ht!]
    \centering
    \includegraphics[width=0.48\textwidth]{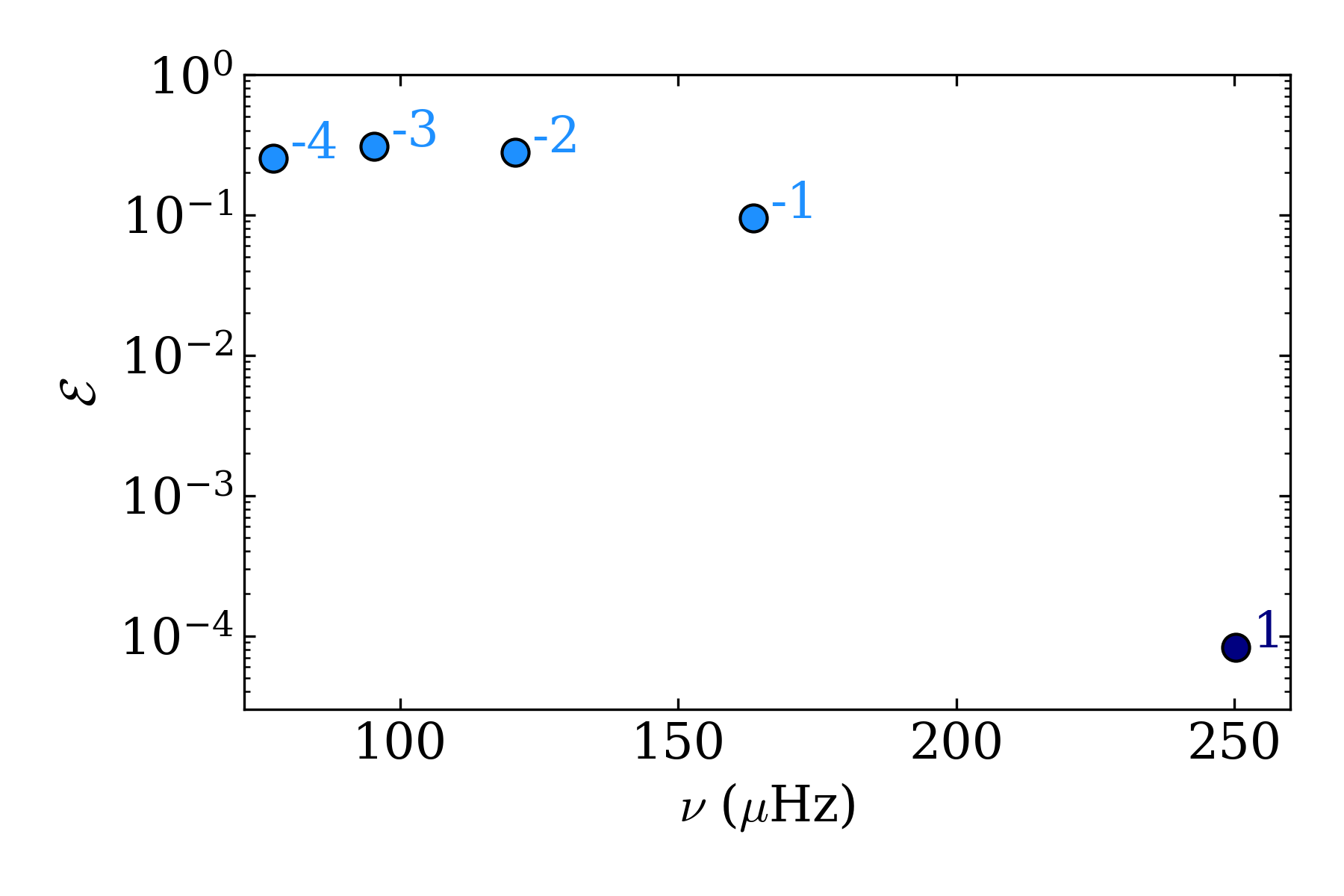}
    \caption{Normalised inertia $\mathcal{E}$ for the mode computed with GYRE with the KIC~3733735 reference model. The g modes are shown in light blue and the p mode in dark blue. The corresponding radial order $n$ is indicated for each mode.
    } 
    \label{fig:inertia_model3733735}
\end{figure}

\begin{figure}[ht!]
    \centering
    \includegraphics[width=0.48\textwidth]{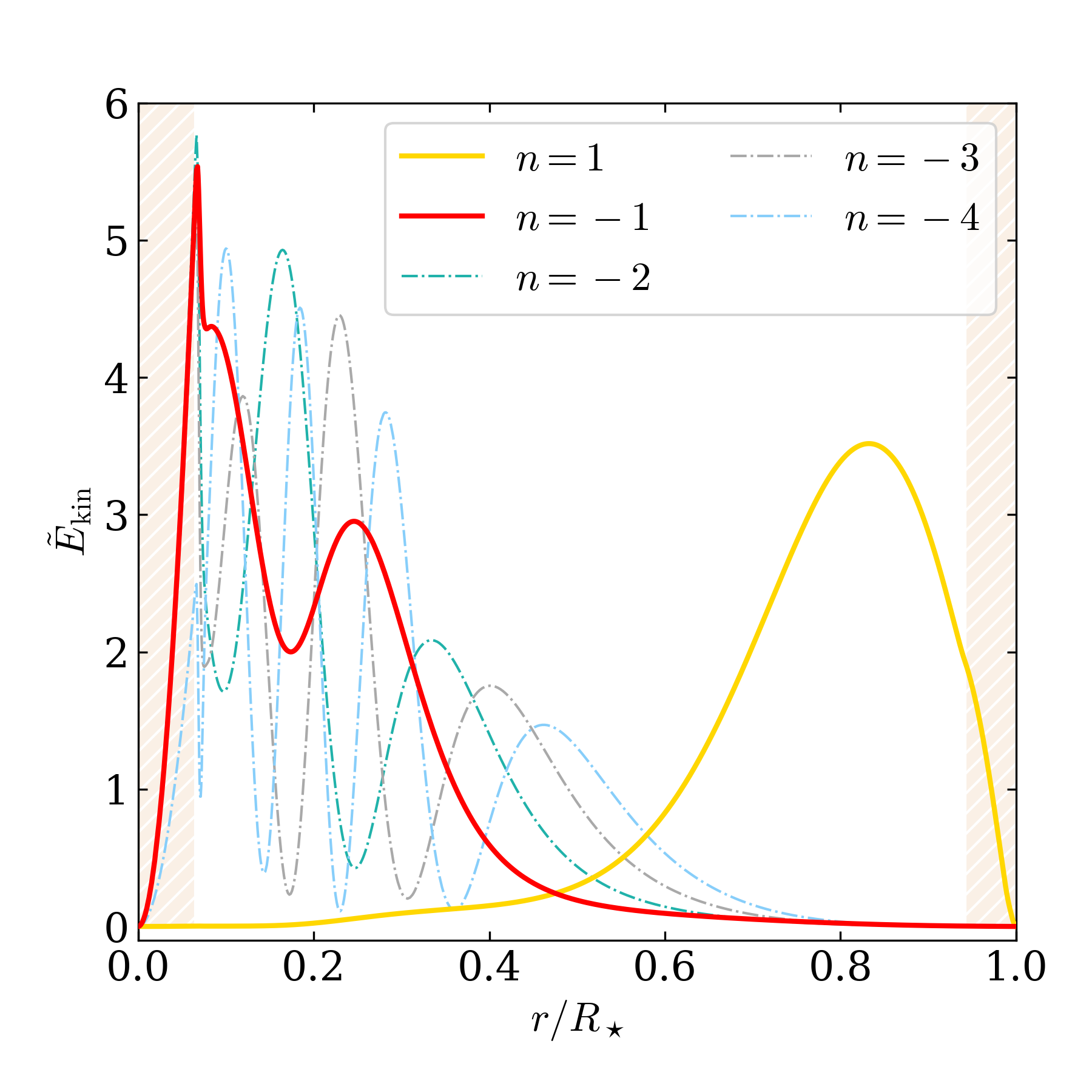}
    \caption{
Normalised kinetic energy density profile $\tilde{E}_\mathrm{kin}$ profiles for the $\ell=1$, $m=0$, $n=1$  (yellow), $-1$ (red), $-2$ (cyan), $-3$ (grey), and $-4$ (blue) modes computed with GYRE for our KIC~3733735 reference model. 
The hatched areas correspond to convective regions.
The $n \leq -2$ modes are shown by the dash-dotted lines for readability. 
   } 
    \label{fig:ked_model3733735}
\end{figure}

The normalised inertia of an oscillation mode is computed as \citep[e.g.][]{ChristensenDaalsgardLectureNotes}
\begin{equation}
    \mathcal{E} = \frac{4 \pi \int_0^{R_\star} [|\xi_r (r)|^2 + \ell (\ell+1) |\xi_h (r)|^2] \rho r^2 dr}{M_\star [|\xi_r (R_\star)|^2 + \ell (\ell+1) |\xi_h (R_\star)|^2] } \; ,
\end{equation}
where $\xi_r$ and $\xi_h$ are the mode vertical and horizontal displacements, respectively. Assuming equipartition of energy, modes with lower normalised inertia will have a larger surface amplitude.
In Fig.~\ref{fig:inertia_model3733735}, we compare $\mathcal{E}$ for the g and p modes shown in Fig.~\ref{fig:3733735_freq_and_period}. The normalised inertia for the $n=1$ p mode is significantly lower than that of the g modes because the resonant cavity of this mode is located closer to the surface. To illustrate this further,  in Fig.~\ref{fig:ked_model3733735} we show the normalised kinetic energy density profile, $\tilde{E}_\mathrm{kin}$, of the modes, which is computed as
\begin{equation}
    \tilde{E}_\mathrm{kin} (r) = \frac{\rho r^2 [|\xi_r (r)|^2 + \ell (\ell + 1)|\xi_h (r)|^2]}{\int_0^1 \rho r^2 [|\xi_r (r)|^2 + \ell (\ell + 1)|\xi_h (r)|^2] \mathrm{d}x} \; ,
\end{equation}
where $x = r/R_\star$ is the dimensionless radius. The normalisation factor $\int_0^1 \rho r^2 [|\xi_r (r)|^2 + \ell (\ell + 1)|\xi_h (r)|^2] \mathrm{d}x$ is chosen such that the integral over $x$ is unity. It appears that the g and p cavities are completely decoupled: for g modes, the kinetic energy density distribution is concentrated in the inner regions of the stars, while the opposite phenomenon is seen for the $n = 1$ p mode. 
For g modes, it can be noted that the behaviour of the mode changes when it enters the convective core. The $\tilde{E}_\mathrm{kin}$ profile suggests that the evanescence of these non-asymptotic g modes in the core is much more limited than in the $n \gg 1$ case. This is also related to the fact that the age yielded by our reference model, namely \num{4.2e8} years, suggests that KIC~3733735 is a young star and the barrier of potential formed by the chemical gradient contribution to $N$ is still limited in this case, allowing g modes to penetrate deeper into the core.
Finally, while we discuss the possible physical explanations for such a behaviour  in Sect.~\ref{sec:inertia_and_power_injection}, we underline here that the g modes and the p mode appear in the S/N spectrum with similar amplitude, while their inertia is different by several orders of magnitude. This suggests that the power injection from the convection to the modes is a strongly frequency-dependent function and increases as frequency decreases.

In the following section, we discuss the case where, contrary to the predictions of the reference model of KIC~3733735, the g- and p-mode resonant cavities of stars are coupled. This latter phenomenon is responsible for the existence of mixed modes with probing abilities both in deep radiative regions and external layers. 

\section{Coupled cavities: Exploring the mixed mode hypothesis \label{sec:mixed_modes}}

In this section, we study the cases of three stars of our sample  in more detail. Indeed, KIC~6679371, KIC~7103006, and KIC~9206432 exhibit a pattern of peaks that can be interpreted as the manifestation of mixed gravito-acoustic modes.

\subsection{Expected properties of mixed modes}

Over a large extent of the frequency range we are probing in this analysis, we expect to have $S_\ell < N$  at some depth of the stellar interior. 
Due to the respective profiles of $N$ and $S_\ell$, the cavities of some low-order g modes should be able to couple with the low-order p-mode resonant cavities through the phenomenon of avoided crossing \citep[see e.g.][]{ChristensenDaalsgardLectureNotes}, where g and p modes exchange properties in their respective resonant cavities when their frequencies are sufficiently close (but never actually cross, hence the name). 
This means that modes with characteristic frequencies close to the eigenfrequencies of both p- and g-mode resonant cavities may have a small normalised inertia while probing the properties of the deepest layers of the star.
As noted for example by \citet{Ong2020}, characterising the behaviour of mixed modes close to an avoided crossing is a very powerful probe of the evolutionary state of the star. 
Indeed, it should be emphasised that, considering the structural evolution of the cavities over time, the efficiency of the coupling may significantly change as the star goes through the MS. While this suggests that only a fraction of stars might exhibit such mixed modes, their detection would provide precious insight into the target where they are observed. 

In MS stars, we expect to see a coupling between the non-asymptotic modes of both cavities. By comparison, in red giant stars, the asymptotic modes of both cavities are coupled, while in subgiant stars, non-asymptotic g modes are coupled to asymptotic p modes \citep[e.g.][]{Ong2020}. 
Contrary to these two latter cases, we therefore expect to have a sparse distribution of both g- and p-dominated modes in the MS targets we are studying.
The properties of the modes within this interval have been explored in the solar case by for example \citet{Provost2000}, but lacking observations, little attention has been paid to this configuration until now. 
\citet{Ong2020} for example focused on evolved solar oscillators, while \citet{Kosovichev2020} and \citet{Bellinger2021} studied the probing capabilities of mixed modes in young subgiant stars. 
Nevertheless, for any $n$, we expect to observe a drop in $\mathcal{E}$ for g-dominated mixed modes compared to pure g modes. 
In what follows, we compare observations to the pattern of $\ell=1$, $m=0$ modes computed with GYRE in an attempt to identify the composition of the patterns we are observing. In addition to the uncertainties introduced by the exact buoyancy profile at the core interface, we expect our computation to be extremely sensitive to the coupling between p- and g- cavities.

We adopt the following nomenclature in what follows. Using GYRE computations, modes with nodes located only in one of the two cavities are referred as pure g or p modes while modes with nodes in both cavities are referred as p- or g-dominated mixed modes, depending on their radial order~$n$. We show in what follows that the mode identification scheme implemented by GYRE \citep{Takata2006} provides surprising results in some cases.

\subsection{KIC~6679371}

\begin{figure}[ht!]
    \centering
    \includegraphics[width=0.48\textwidth]{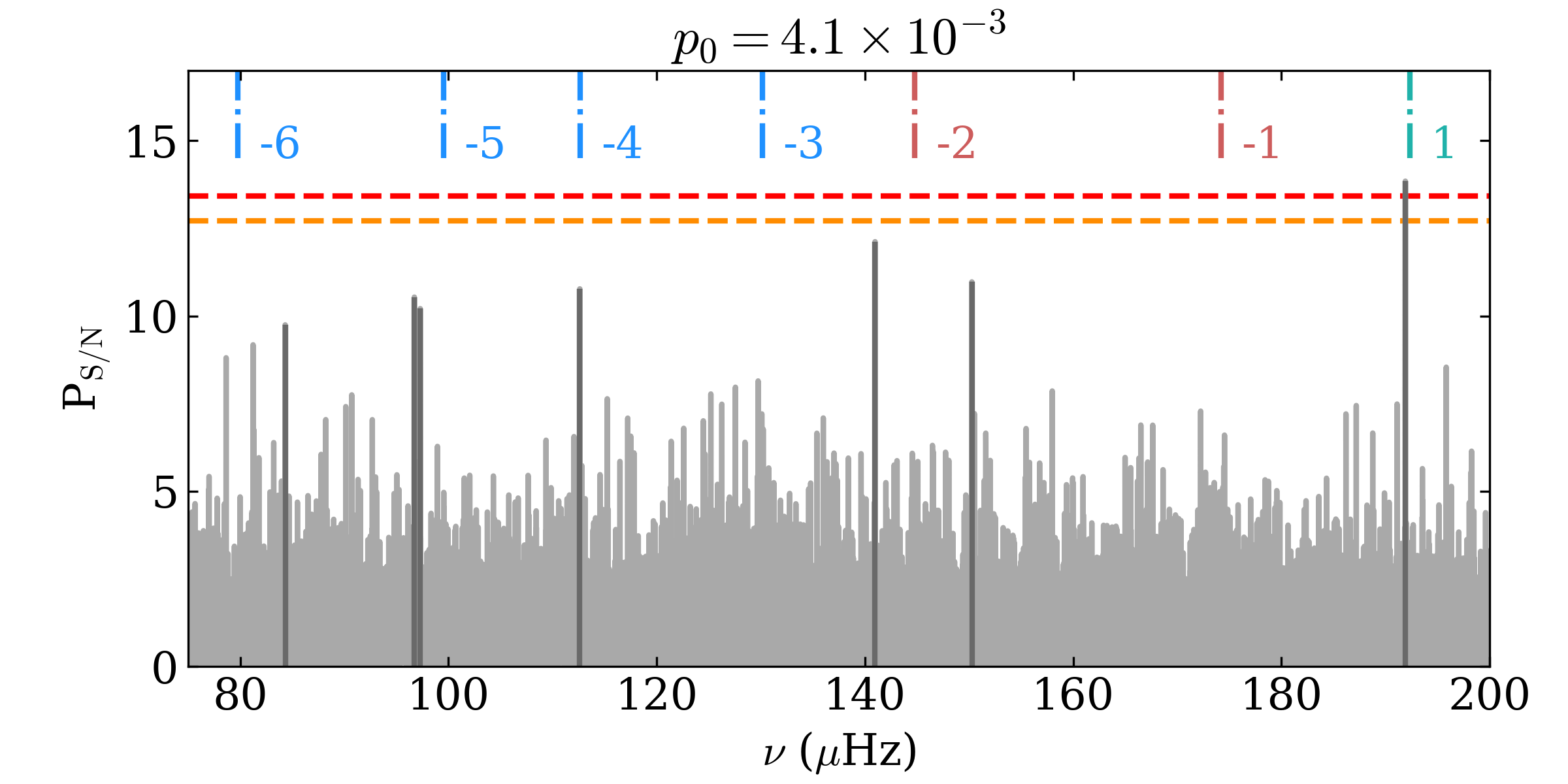}
    \includegraphics[width=0.48\textwidth]{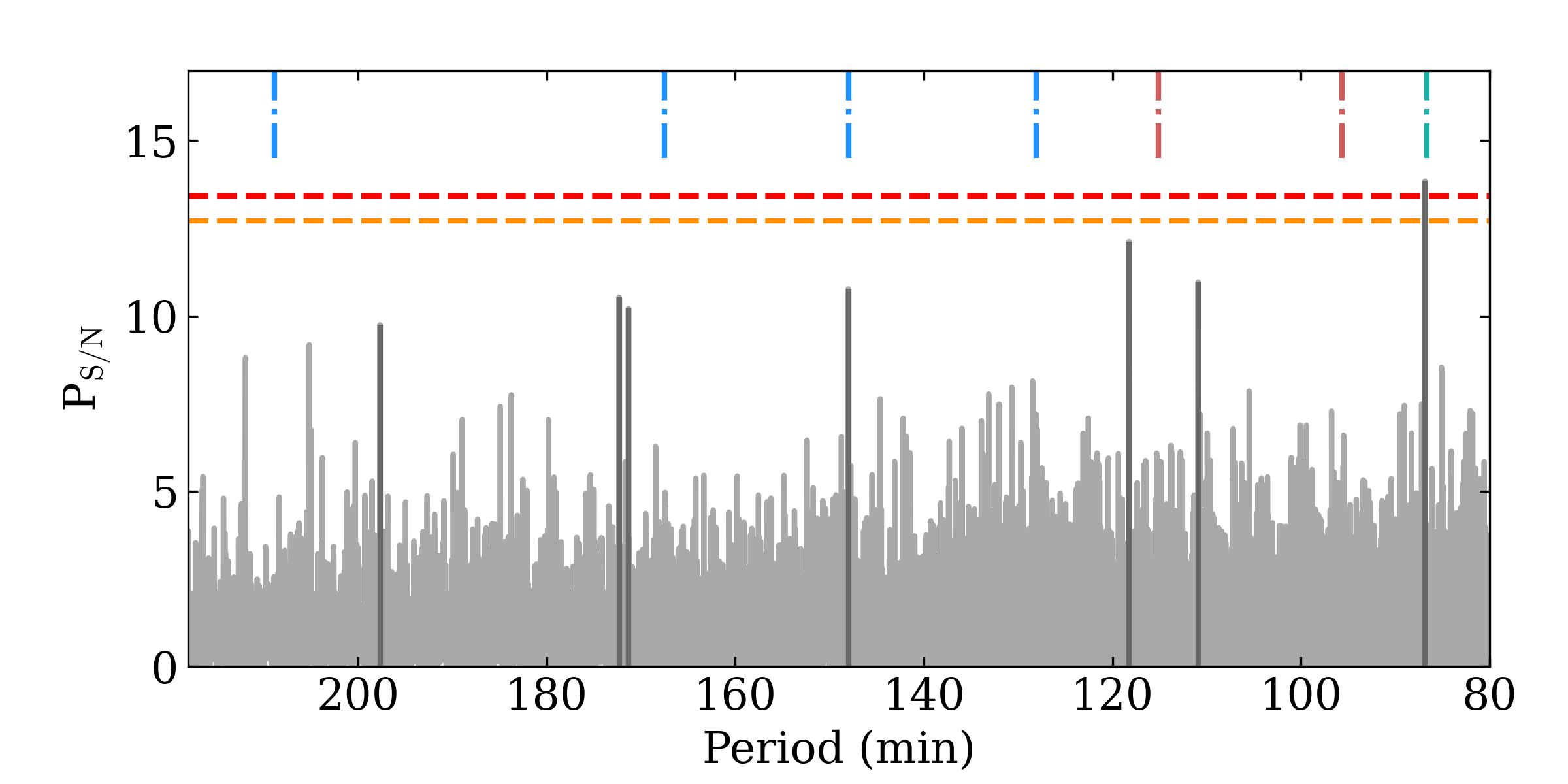}
    \caption{
    $\mathrm{P}_\mathrm{S/N}$ (\textit{grey}) obtained for KIC~6679371, with the x-axis scaled in frequency (\textit{top}) and in period (\textit{bottom}). 
    The dashed horizontal orange and red lines correspond to the $p_\mathrm{det} = 0.1$ and 0.05 detection thresholds computed from Eq.~(\ref{eq:appourchaux2000}), respectively. The peaks included when computing the pattern probability $p_0$ are shown in darker grey.
    The mode frequencies computed with GYRE for $\ell=1$, $m=0$ modes are shown for  pure g modes (light blue), g-dominated mixed modes (red), and g-dominated mixed modes (green). The order $n$ of each modes is shown in the figure.
    } 
    \label{fig:6679371_freq_and_period}
\end{figure}

The first star with potential mixed modes we consider is KIC~6679371.
In Fig.~\ref{fig:6679371_freq_and_period}, we show the pattern we analyse and compare it with the frequency computed with our reference model. The noise probability of the selected pattern is $p_0 = \num{4.1e-3}$. The highest amplitude peak at 191.9~$\mu$Hz is detected by the homogeneous analysis presented in Sect.~\ref{sec:ensemble_analysis} and is above the $p_\mathrm{det}=0.05$ detection threshold. We match it to the $n=1$ p-dominated mixed mode of the reference model. No observed peak can be matched to the $n=-1$ g-dominated mixed mode, while the remaining peaks could correspond to a pattern of modes with orders between  $n=-2$ and $n=-6$, although we have a discrepancy between the observed and modelled frequencies in some cases. The predicted frequencies of the $n=-4$ and $n=-5$ modes are nevertheless in good agreement with the observed peaks.

\begin{figure}[ht!]
    \centering
    \includegraphics[width=0.48\textwidth]{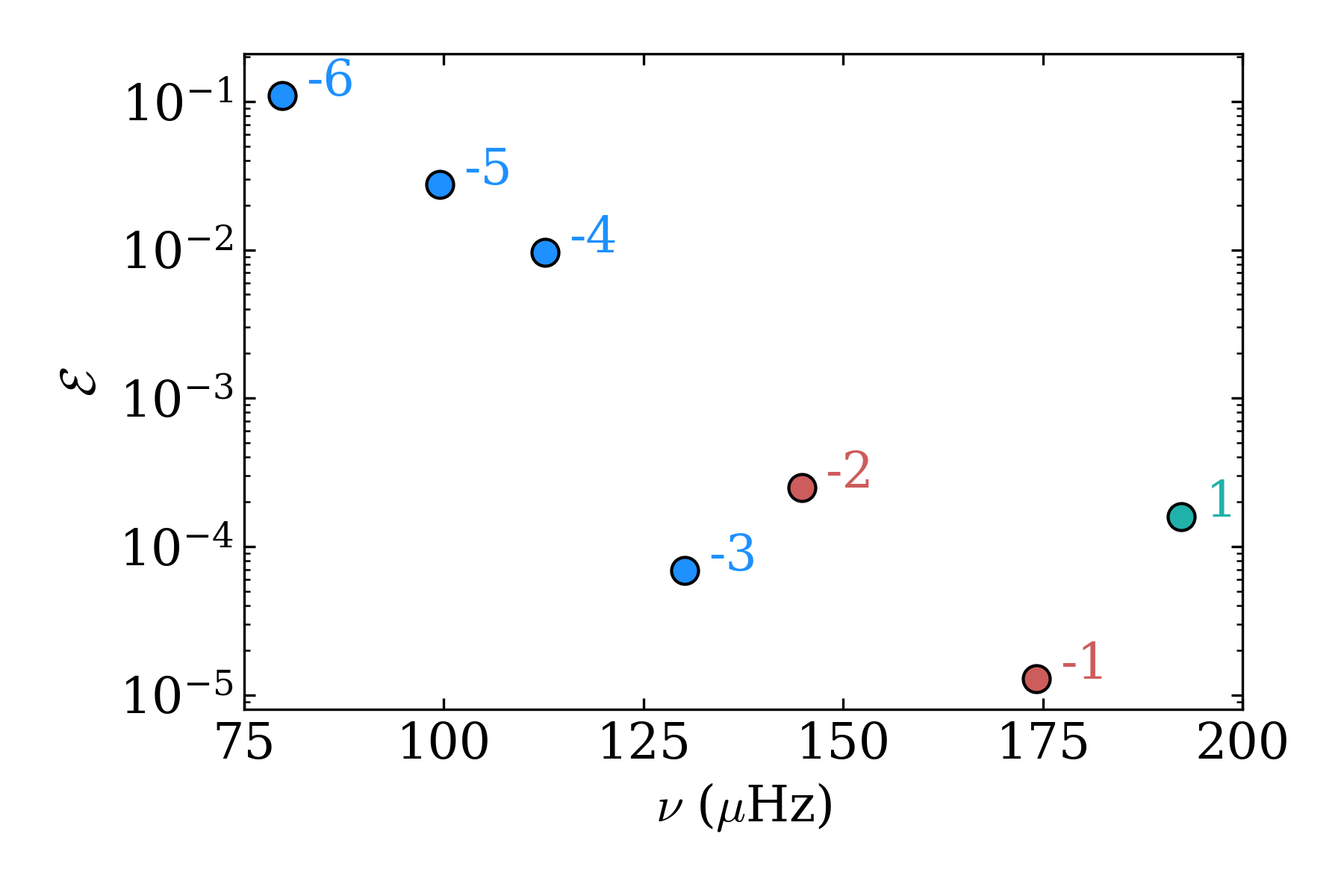}
    \caption{Normalised inertia $\mathcal{E}$ for the modes computed with GYRE with the KIC~6679371 reference model. The pure g modes are shown in light blue, the g-dominated mixed modes in red, and the p-dominated mixed mode in green. The corresponding radial order $n$ is indicated for each mode.
    } 
    \label{fig:inertia_model6679371}
\end{figure}

\begin{figure}[ht!]
    \centering
    \includegraphics[width=0.48\textwidth]{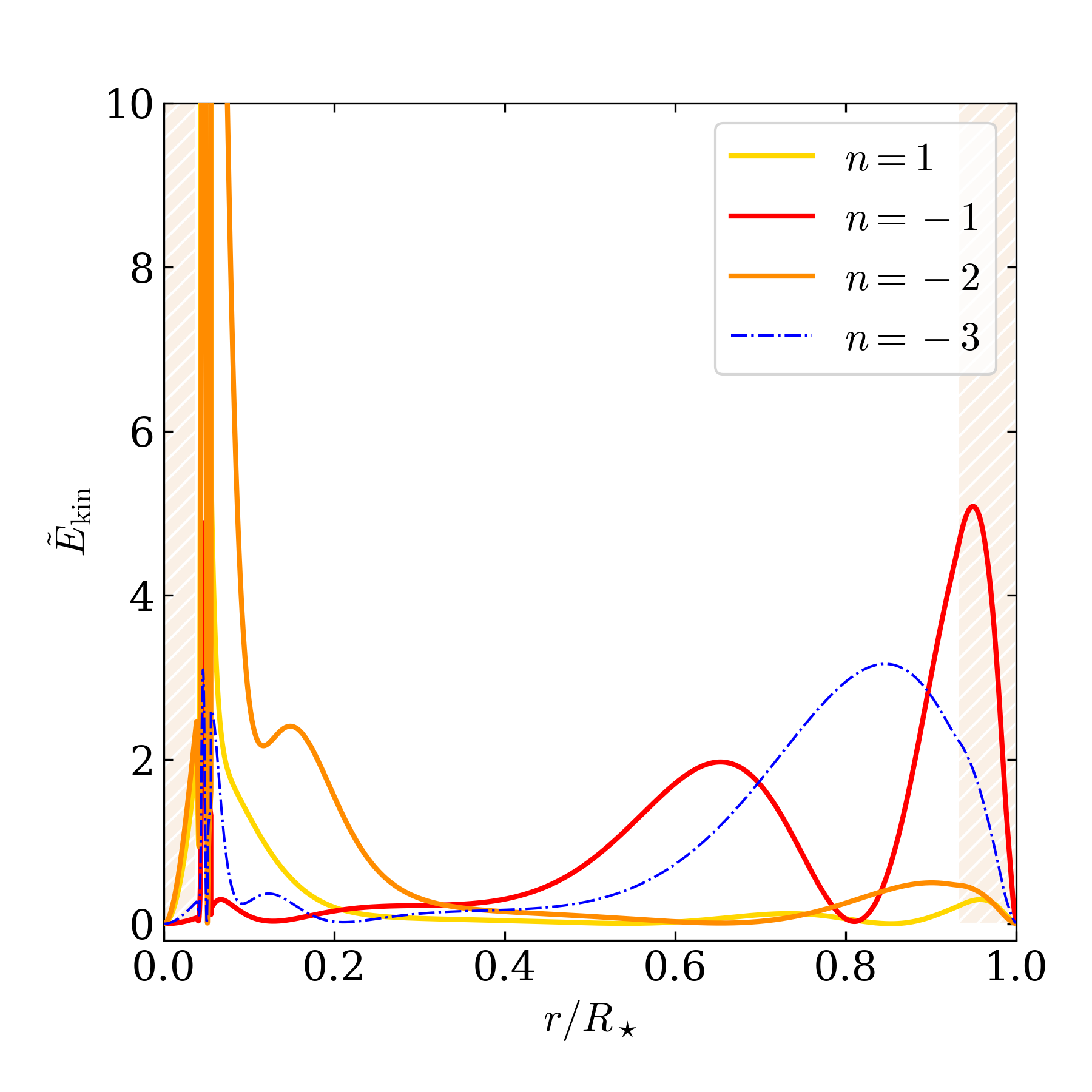}
    \caption{
    Normalised kinetic energy density $\tilde{E}_\mathrm{kin}$ profiles for mixed modes computed with GYRE for our KIC~6679371 reference model. The dash-dotted blue line shows the $\tilde{E}_\mathrm{kin}$ profile for the $n = -3$ mode, which has no node in the p-mode cavity according to GYRE but exhibits a quasi-mixed mode behaviour.
    The hatched areas correspond to convective regions.
    } 
    \label{fig:ked_model6679371}
\end{figure}

The normalised mode inertia values are represented in Fig.~\ref{fig:inertia_model6679371}. As expected, the g-dominated mixed modes have significantly lower inertia than the pure g modes. The case of the $n=-3$ mode is interesting as it is labelled as a pure g mode and has no node in the p-mode cavity according to GYRE. When we represent the kinetic energy density in Fig.~\ref{fig:ked_model6679371}, we observe that, for
this mode, a significant fraction of the kinetic energy is located in the upper layer of the star, which is not expected for a pure g mode. We suggest that, even if the mode has no p-mode cavity node, its own g-mode resonant cavity is sufficiently affected by the proximity of the p-mode resonant cavity that it is a quasi-mixed mode.  
We also highlight that the $n=1$ p-dominated mixed mode computed by GYRE has an $\tilde{E}_\mathrm{kin}$ profile peaked at the bottom of the radiative zone, while we would have expected it to be located much closer to the surface. Consequently, it has a lower inertia than the $n=-1$ g-dominated mode (which we do not identify in the observed pattern). We believe that this behaviour is strongly related to the profile of $N$ close to the convective core interface, meaning that the mode is sensitive to the chemical-composition gradient at the interface. 
We remind the reader that asymptotic modes are less affected by the exact profile of the resonant cavities boundary, which allows us to study their properties using the Wentzel-Kramers-Brillouin-Jeffreys \citep[WKBJ,][]{Froman1965} approximation. 
We suggest that a careful examination of the impact of stellar modelling choices \citep[especially concerning the way core overshooting is accounted for; e.g.][]{Deheuvels2016} would allow a better understanding of the behaviour of non-asymptotic modes but constitutes an independent work that is out of the scope of this paper. Nevertheless, this suggests that such non-asymptotic modes could be used to probe overshoot and induced mixing at the convective core interface \citep[see e.g.][]{Mombarg2019,Mombarg2021,Aerts2021b}.

\subsection{KIC~7103006}

\begin{figure}[ht!]
    \centering
    \includegraphics[width=0.48\textwidth]{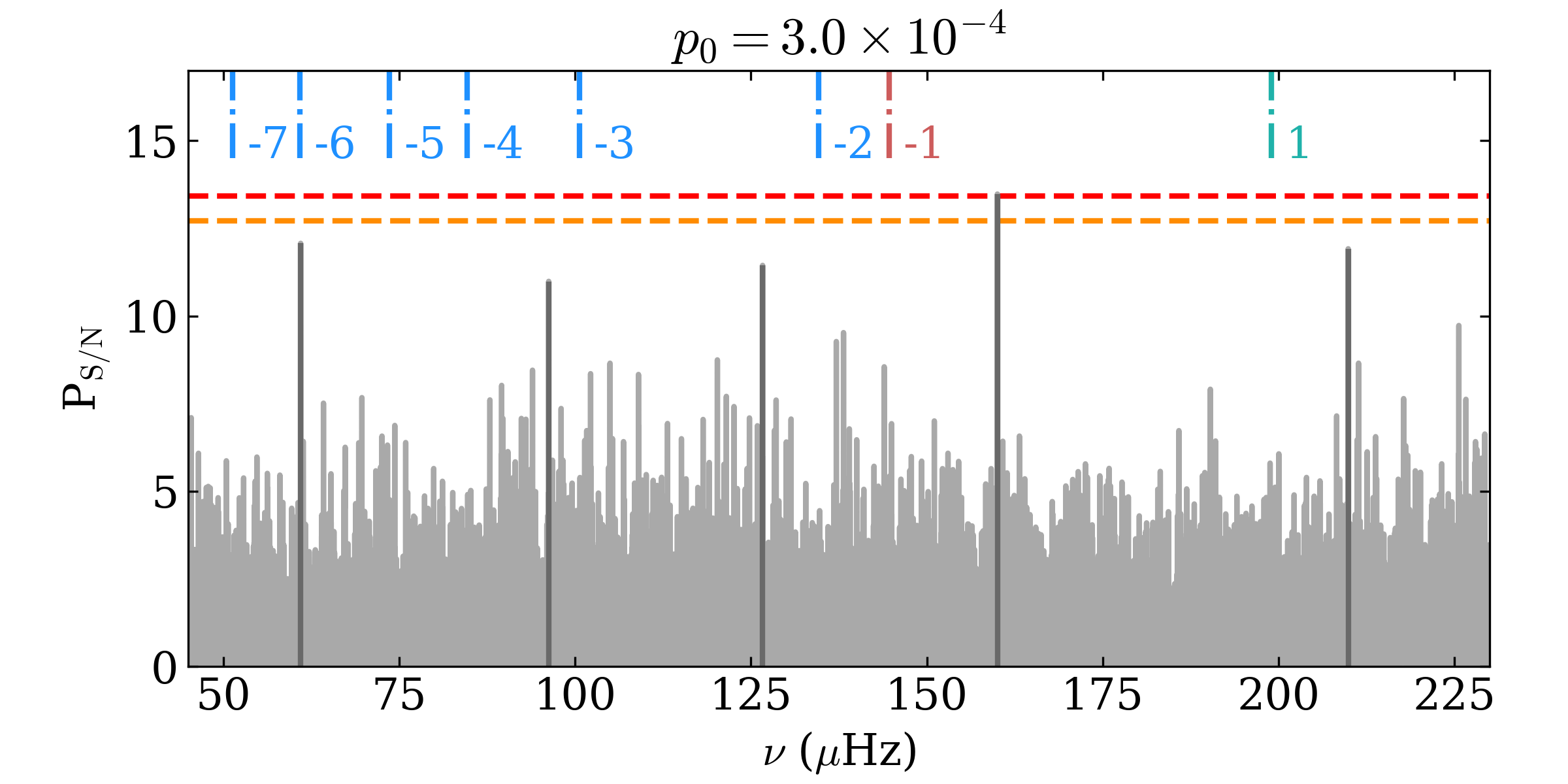}
    \includegraphics[width=0.48\textwidth]{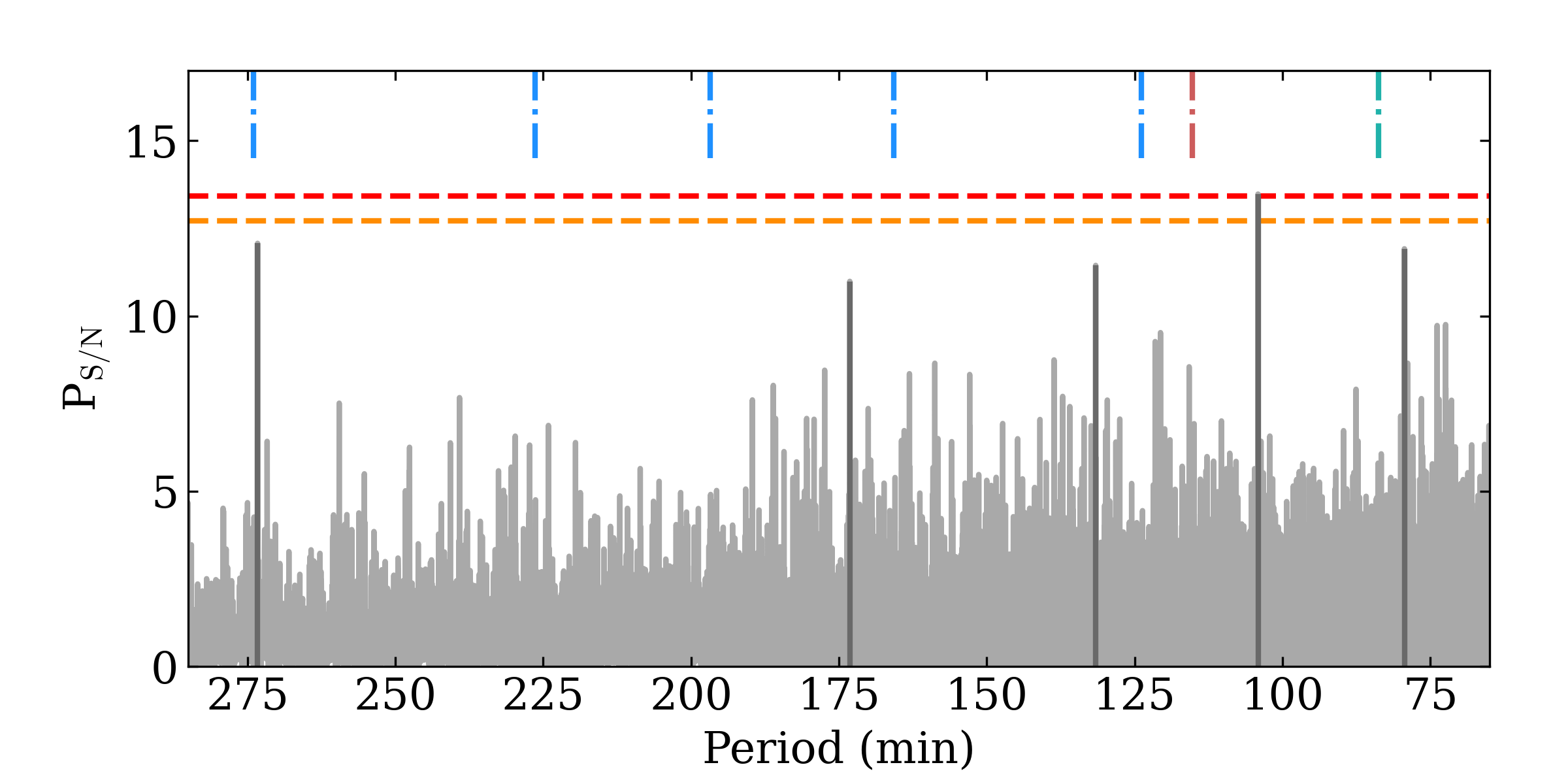}
    \caption{
    Same as Fig.~\ref{fig:6679371_freq_and_period} but for KIC~7103006.
    } 
    \label{fig:7103006_freq_and_period}
\end{figure}

\begin{figure}[ht!]
    \centering
    \includegraphics[width=0.48\textwidth]{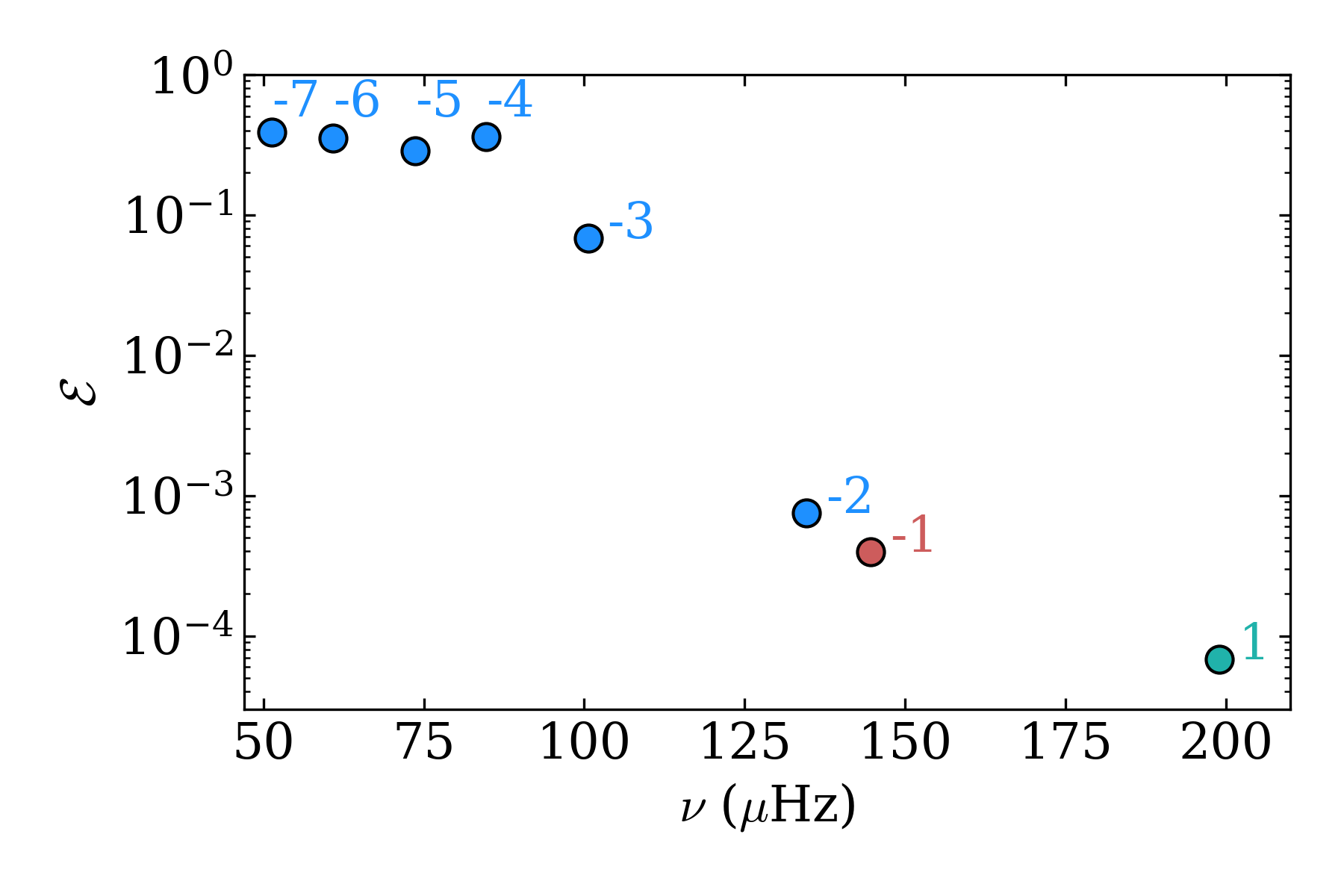}
    \caption{
    Same as Fig.~\ref{fig:inertia_model6679371} but for KIC~7103006.
    } 
    \label{fig:inertia_model7103006}
\end{figure}

\begin{figure}[ht!]
    \centering
    \includegraphics[width=0.48\textwidth]{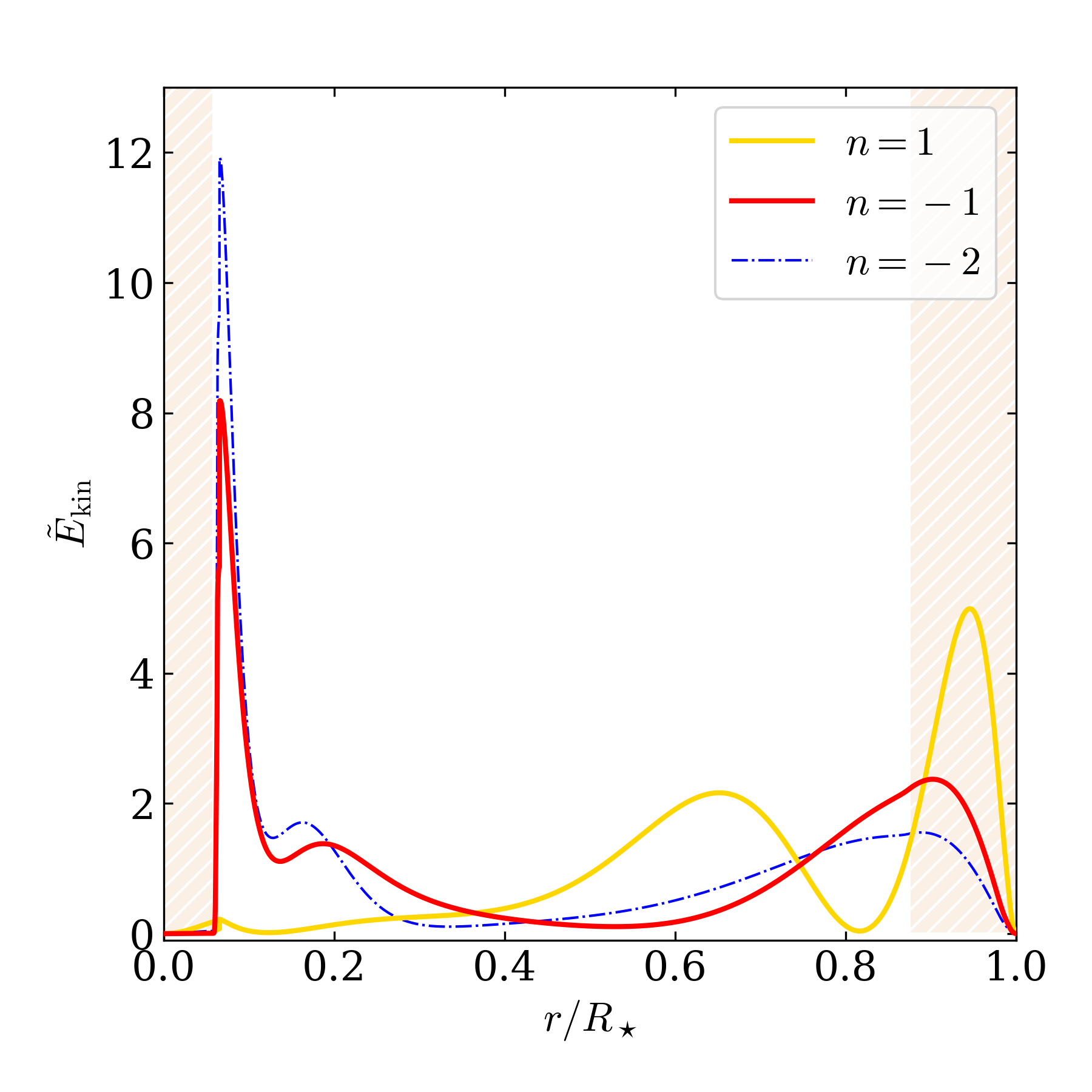}
    \caption{
    Normalised kinetic energy density $\tilde{E}_\mathrm{kin}$ profiles for mixed modes computed with GYRE for our KIC~7103006 reference model. The dash-dotted blue line shows the $\tilde{E}_\mathrm{kin}$ profile for the $n = -2$ mode, which has no node in the p-mode cavity according to GYRE but has a $\tilde{E}_\mathrm{kin}$ similar to that of the $n=-1$ and $n=1$ mixed modes.
    } 
    \label{fig:ked_model7103006}
\end{figure}

The second star on which we identify a possible pattern is KIC~7103006, with $p_0 = \num{3.0e-4}$. As we can see in Fig.~\ref{fig:7103006_freq_and_period}, in this case we are also able to match the peak at highest frequency (209.9~$\mu$Hz) to the $n=1$ p-dominated mixed mode of the reference model, although with a larger frequency discrepancy than the previous case. The 160.0~$\mu$Hz peak detected by the analysis from Sect.~\ref{sec:ensemble_analysis} could correspond to the $n=-1$ g-dominated mode. It is more difficult to proceed to a formal identification for the following peaks, which are located in an area where GYRE predicts pure g modes.  Figure~\ref{fig:inertia_model7103006} shows the normalised inertia of the modes where we are again able to verify that the normalised inertia of the g modes increases as frequency decreases. Figure~\ref{fig:ked_model7103006} shows the kinetic energy density of the $n=-2$, $-1,$ and $1$ modes. As in the previous case, the $n=-2$ mode does not have any node in the p-mode cavity but exhibits a quasi-mixed behaviour very similar to that of the $n=-1$ g-dominated mixed mode. The sensitivity of the $n=1$ p-dominated mixed modes is concentrated in the upper half of the star.

\subsection{KIC~9206432}

\begin{figure}[ht!]
    \centering
    \includegraphics[width=0.48\textwidth]{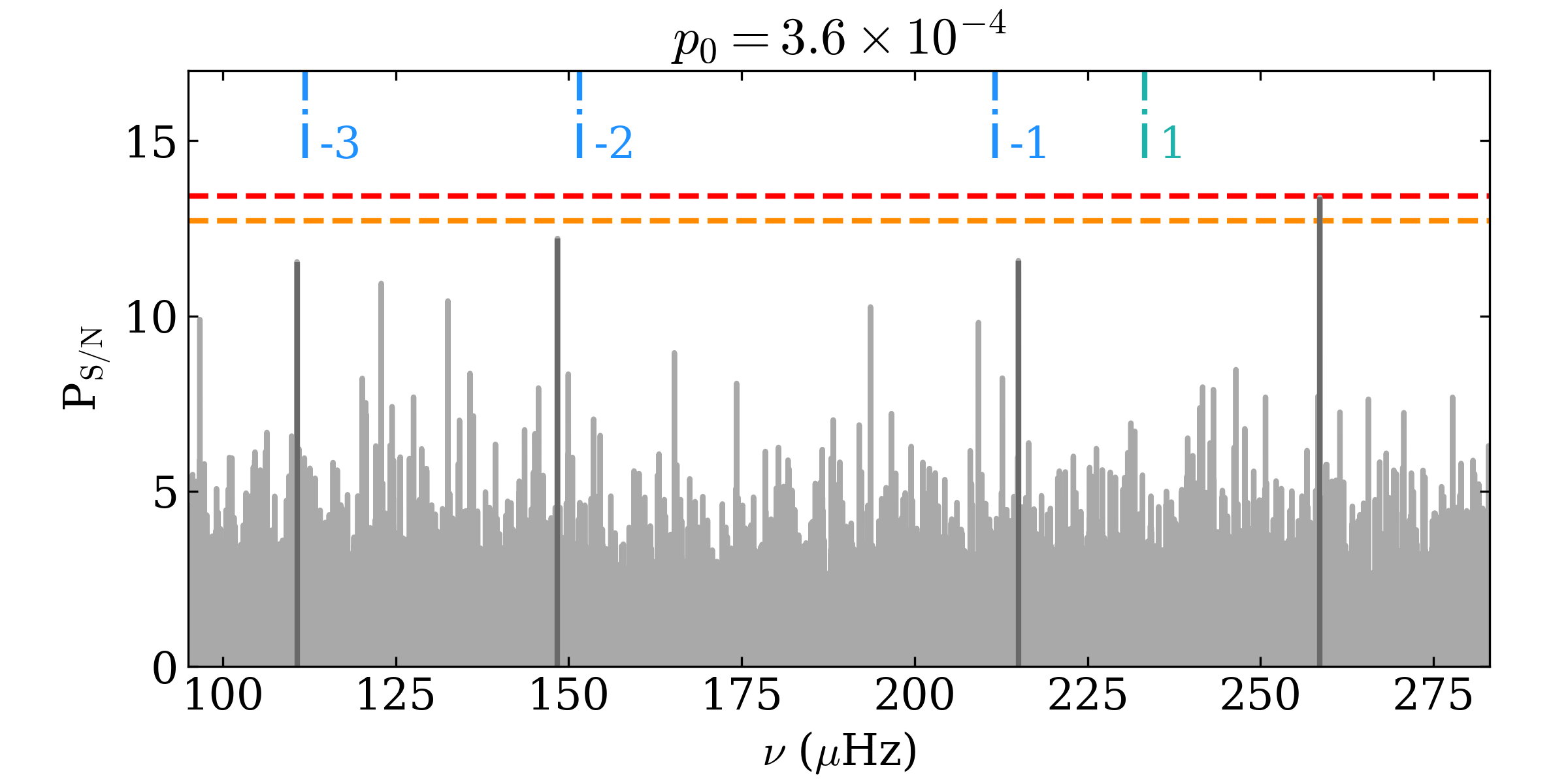}
    \includegraphics[width=0.48\textwidth]{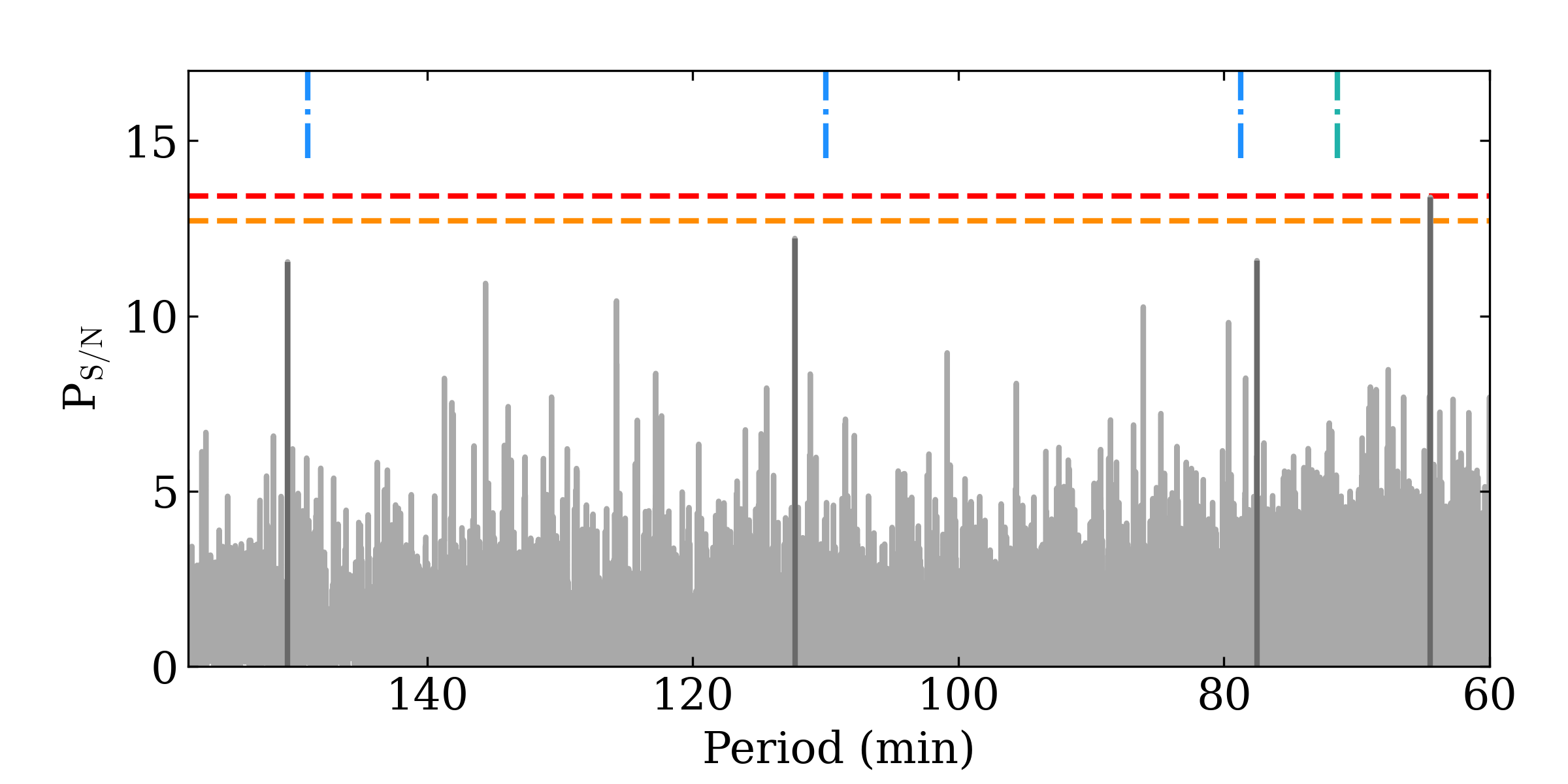}
    \caption{
    Same as Fig.~\ref{fig:6679371_freq_and_period} but for KIC~9206432.
    } 
    \label{fig:9206432_freq_and_period}
\end{figure}

Finally, we take a look at KIC~9206432, where the pattern we select has $p_0 = \num{3.6e-4}$.
As visible in Fig.~\ref{fig:9206432_freq_and_period}, the highest-amplitude peak is located at 258.6~$\mu$Hz, which is relatively far from the $n=1$ p-dominated mixed modes predicted by GYRE at 233.3~$\mu$Hz, which could be explained by a significant difference in the way cavities are coupled in the reference model with respect to the actual $N$ profile. We match the three other detected peaks to the $n=-3$, $-2,$ and $-1$ pure g modes of the reference model, all with good agreement between the GYRE predicted frequencies and the position of the detected peaks.

\begin{figure}[ht!]
    \centering
    \includegraphics[width=0.48\textwidth]{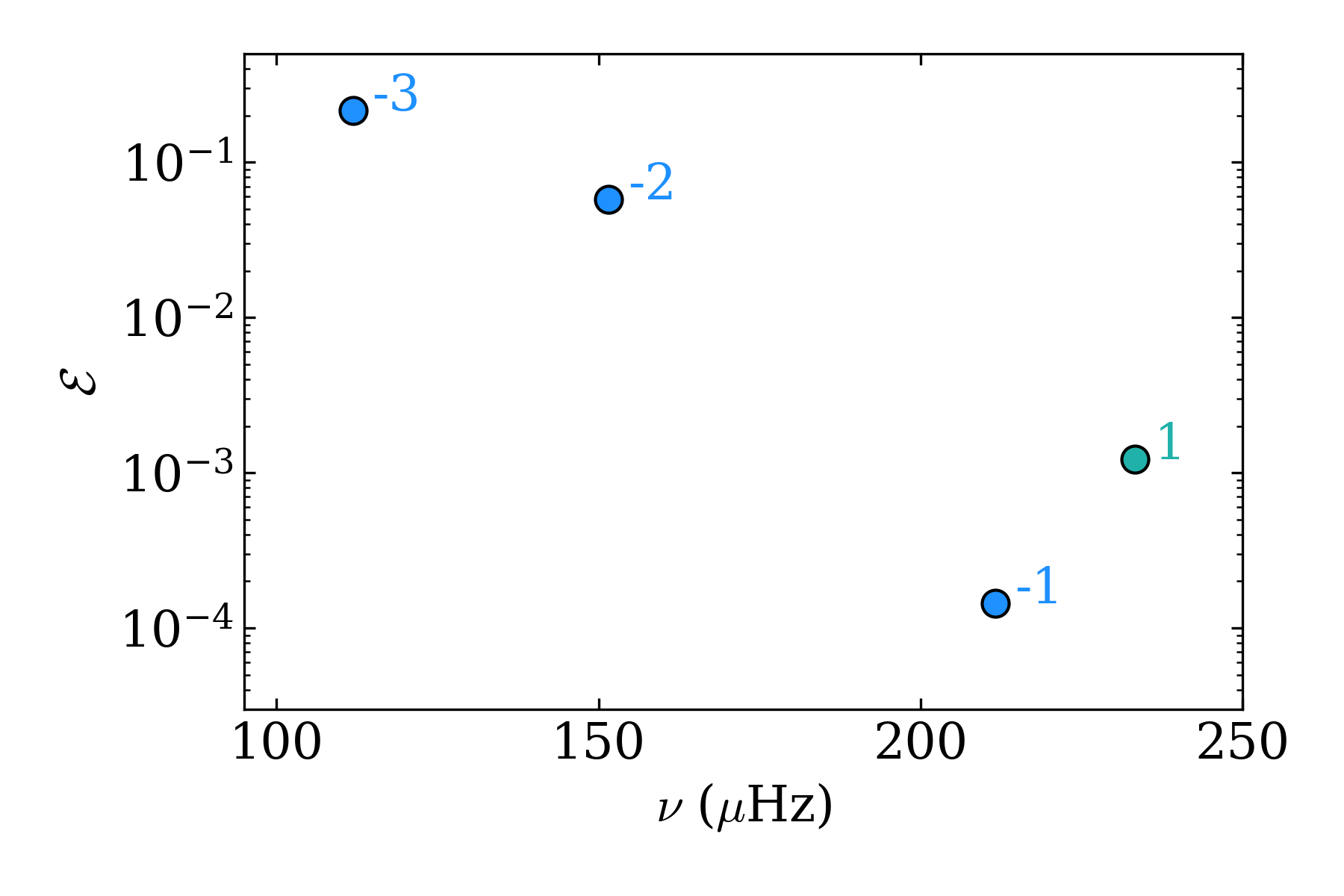}
    \caption{
    Same as Fig.~\ref{fig:inertia_model6679371} but for KIC~9206432.
    } 
    \label{fig:inertia_model9206432}
\end{figure}

\begin{figure}[ht!]
    \centering
    \includegraphics[width=0.48\textwidth]{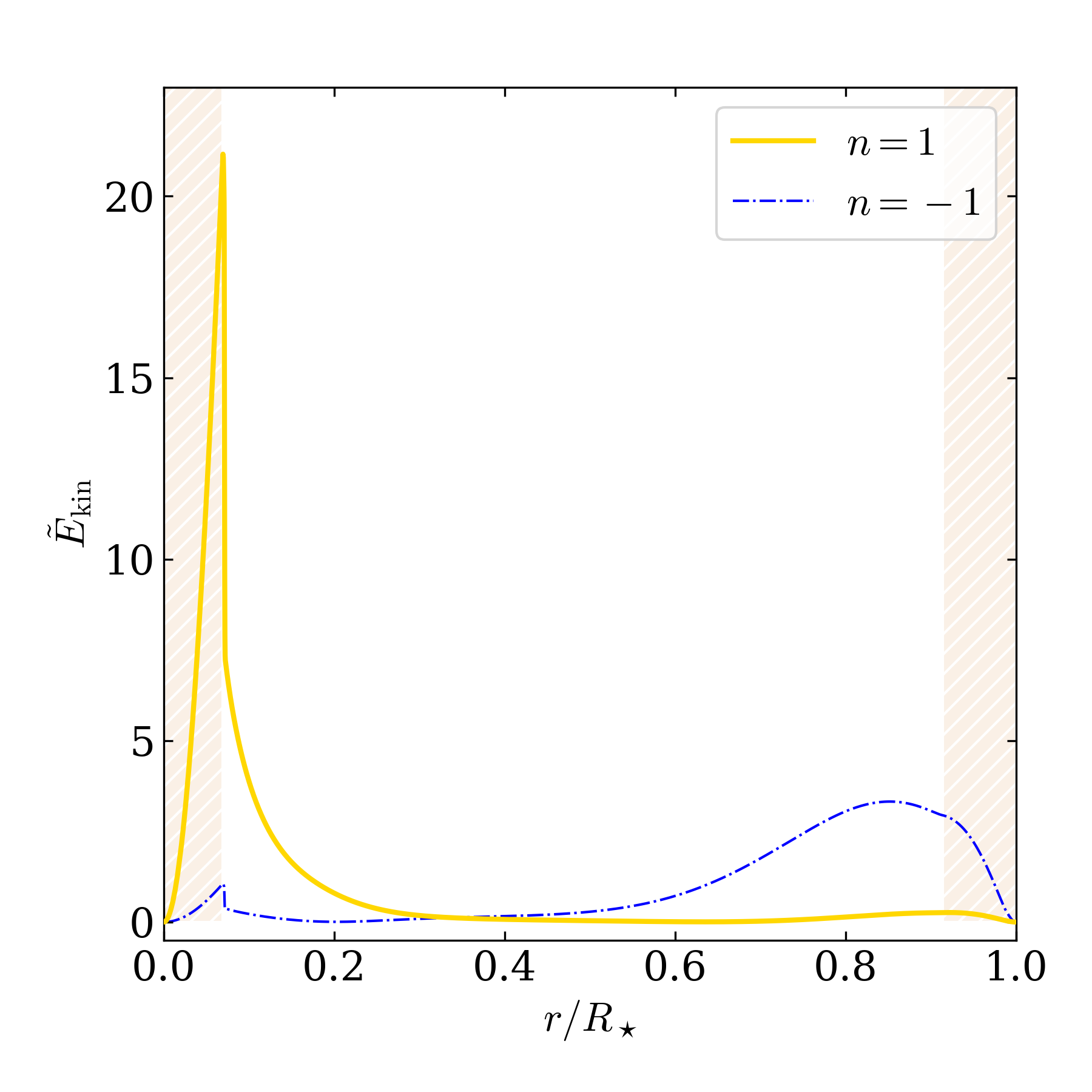}
    \caption{
    Normalised kinetic energy density $\tilde{E}_\mathrm{kin}$ profiles for mixed modes computed with GYRE for our KIC~9206432 reference model. The dash-dotted blue line shows the $\tilde{E}_\mathrm{kin}$ profile for the $n = -1$ mode, which has no node in the p-mode cavity according to GYRE but has a $\tilde{E}_\mathrm{kin}$ similar to the $n=-1$ and $n=1$ mixed modes.
    } 
    \label{fig:ked_model9206432}
\end{figure}

The corresponding mode inertia are shown in Fig.~\ref{fig:inertia_model9206432}, where we can see that $n=-1$ and $1$ modes have a lower computed inertia than their lower-frequency counterparts. The $\tilde{E}_\mathrm{kin}$ profile obtained for the $n=-1$ and $1$ modes (Fig.~\ref{fig:ked_model9206432}) is puzzling (close to what we observed in the case of KIC~6679371). Indeed, the $n=1$ p-dominated mixed mode energy density is located close to the convective core interface, while the $n=-1$ pure g mode kinetic energy density is distributed much higher in the star, whilst at the same time  having no node in the p-mode cavity.

\subsection{Mode inertia and power injection \label{sec:inertia_and_power_injection}}

We finally emphasise that, with regards to the similar S/Ns observed for the peaks of the patterns discussed above, the difference in inertia predicted by GYRE between pure g~modes and mixed modes suggests that the mode excitation is more efficient in the frequency range where we identified peaks as pure g~modes than in the frequency interval where the cavity coupling occurs. 
As stated above, a similar behaviour was noted in the case of KIC~3733735, where no cavity coupling is expected.
This is nevertheless consistent with the trend predicted by semi-analytical predictions \citep[e.g.][]{Pincon2016} and numerical simulations \citep[e.g.][]{Rogers2013}, especially when the action of rotation is taken into account. \citet{Augustson2020} indeed predicted that, accounting for rotation, the power injection from convection to the waves could significantly increase in the lowest frequency intervals of the super inertial regime. Additionally,
\citetalias{Breton2022simuFstars} were able to show that rotation significantly affects the shape of the mode power injection function (see e.g. their Fig.~20, where mode amplitude below 50~$\mu$Hz changes by several orders of magnitude as the rotation frequency increases).

\section{Surface velocity \label{sec:upper_threshold}}

Having discussed possible matches from reference models to patterns of interests observed in some stars of our sample, we propose in this section to use our analysed sample to provide an estimate of the equivalent mode velocity related to the significant peaks we detect in Sect.~\ref{sec:stat_analysis} and to derive an upper threshold of the g- and/or mixed-mode surface velocity in this population of stars. 

\subsection{From observed luminosity variation to mode equivalent velocity}

In \citetalias{Breton2022simuFstars}, the root mean square (r.m.s) luminosity perturbation, $\delta L_\star$, related to the action of a single oscillation mode of degree $\ell$ is linked to the mode temporal r.m.s radial velocity $\left<v_r\right>$ through the following relation:
\begin{equation}
\label{delta_L_L_non_integrated}
    \left . \frac{\delta L_\star}{L_\star} \right|_\mathrm{mode} = 
    \Bigg[ 
    4\nabla_\mathrm{ad} \left| \frac{\ell (\ell+1)}{\overline{\omega}^2} - 4 - \overline{\omega}^2 \right| 
    + 2
    \Bigg] \frac{\left<v_r\right>}{\omega R_\star} \; ,
\end{equation}
where $\nabla_\mathrm{ad}$ is the adiabatic gradient. The brackets $<>$ denote the r.m.s value of the corresponding quantity. The reduced frequency $\overline{\omega}$, which is the frequency normalised by the dynamical frequency, is given by
\begin{equation}
    \overline{\omega}^2 = \frac{\omega^2 R_\star^3}{G M_\star} \; ,
\end{equation}
where $G$ is the gravitational constant.
This relation is obtained by considering the adiabatic Lagrangian pressure perturbation at the surface of the star arising from the action of a single mode of degree $\ell$ \citep{Dzembowski1971,Buta1979}, in the \citet{Cowling1941} approximation.
However, this quantity corresponds to a bolometric perturbation, which needs to be integrated on the stellar disc accounting for the spherical harmonic structure of the mode in order to predict the actual luminosity perturbation in disc-integrated observations; it also
does not account for limb-darkening, instrumental bandwidth, or transfer function.

To account for these effects, we consider a spherical system of coordinates $(r, \theta, \phi)$ and we follow the prescription from \citet[][hereafter BP90]{Berthomieu1990} to write the luminosity fluctuation arising from the action of a single mode of degree $\ell$ and azimuthal number $m$ in the adiabatic approximation
\begin{equation}
\begin{aligned}
    \label{eq:deltaL_L}
  \left . \frac{\delta L_\star}{L_\star} (t, \theta_0, \phi_0) \right|_{\ell,m} &= 
  \left< \frac{|\delta r|^2}{R_\star^2} \right>^{1/2} \left( \frac{(2\ell+1)(l-m)!}{4\pi (\ell+m)!} \right)^{1/2} \\
  &\times \frac{1}{b_0}\left[ 4 \nabla_\mathrm{ad} \left ( \frac{\ell (\ell + 1)}{\overline{\omega}^2} - 4 - \overline{\omega}^2 \right) b_\ell + 2 b_\ell - c_\ell \right] \\
  &\times P_\ell^m (\cos \theta_0) \cos (\omega t + m \phi_0)
    \; , 
\end{aligned}
\end{equation}
where $\theta_0$ and $\phi_0$ denote the co-latitude and the longitude of the observer line of sight with respect to the stellar equatorial plane. The function $P_\ell^m$ is the associated Legendre polynomial of degree $\ell$ and azimuthal number $m$.
The relation that we are discussing here is not affected by the nature of the mode, the considered mode can be a pure p mode, a pure g mode, or a mixed mode.
The mean relative displacement $\left<|\delta r|^2 / R_\star^2\right>$ is related to the radial component of the mode eigenfunction $\xi_r$ through 
\begin{equation}
\left< \frac{|\delta r|^2}{R_\star^2} \right> = 2\pi \int_{-\pi}^{\pi} \sin \theta |\xi_r|^2 d\theta \; ,
\end{equation}
and can be connected to $\left<v_r\right>$ through the simple relation
\begin{equation}
\left<v_r\right>^2 = \left< |\delta r|^2\right> \omega^2 \; .
\end{equation}
The integrals $b_\ell$ and $c_\ell$ were defined by \citet{Dziembowski1977} and are written as
\begin{equation}
\label{eq:b_l_c_l}
\begin{aligned}
    &b_\ell (\mu) = \int_0^1 \mu P_\ell (\mu) W (\mu) d\mu \\ 
    &c_\ell (\mu) = \ell \int_0^1 \left(W (\mu) - \mu \diffp{W}{\mu} \right) (P_{\ell-1} (\mu) - \mu P_\ell (\mu)) d\mu
    \; ,
\end{aligned}
\end{equation}
where $\mu = \cos \theta$,
the function $P_\ell$ is the Legendre polynomial of degree $\ell$, and $W$ is a weighting function that depends both on the stellar atmospheric properties and on the instrumental transfer function.
Given the specific intensity $I_K (\mu)$, we compute $W (\mu)$ as the $I_K (\mu) / I_K(1)$ ratio, using the four-coefficient non-linear relation provided for \textit{Kepler} by \citet{Sing2010}.

We can rewrite Eq.~(\ref{eq:deltaL_L}) under the following form: 
\begin{equation}
    \left . \frac{\delta L_\star}{L_\star} (t, \theta_0, \phi_0) \right|_{\ell,m} = 
    \left< v_r \right> A_{\ell, m} (\theta_0) \cos (\omega t + m\phi_0), \;
\end{equation}
where
\begin{equation}
\begin{aligned}
    A_{\ell, m} (\theta_0) &= \frac{1}{\omega R_\star} \left( \frac{(2\ell+1)(l-m)!}{4\pi (\ell+m)!} \right)^{1/2} \\
  &\times \frac{1}{b_0}\left[ 4 \nabla_\mathrm{ad} \left ( \frac{\ell (\ell + 1)}{\overline{\omega}^2} - 4 - \overline{\omega}^2 \right) b_\ell + 2 b_\ell - c_\ell \right] \\
  &\times P_\ell^m (\cos \theta_0) \; ,
\end{aligned}
\end{equation}
and we are able to connect the r.m.s luminosity fluctuation $\left < \delta L_\star / L_\star \right>$ to $\left< v_r \right>$ by writing 
\begin{equation}
\label{eq:deltaL_L_rms}
  \left. \left < \frac{\delta L_\star}{L_\star} \right> (\theta_0) \right|_{\ell,m} = \left< v_r \right> | A_{\ell, m} (\theta_0) | \; .
\end{equation}

This formula is valid as long as the stellar rotational frequency is small with regards to the oscillation frequency. For subinertial modes, that is modes with angular frequency below $2 \Omega_\star$, a modified formalism has to be applied \citep[see][]{Townsend2003b}. However, the subinertial frequency range of the  PSDs of solar-type stars hosts complex quasi-harmonic patterns related to surface rotation (through the apparition and evolution of active regions in the photosphere). Some of these modulations might also possibly be related to the modulations from r modes \citep[e.g.][]{Loptien2018,Saio2018,Gizon2021} or thermal Rossby waves \citep[e.g.][]{Hindman2022}. Deriving an upper amplitude threshold for gravito-inertial modes in this frequency range is therefore beyond the scope of this paper.
It should also be noted that the accuracy of the luminosity fluctuation defined by Eq.~(\ref{eq:deltaL_L}) relies on the adiabatic approximation we made. We discuss the limitations of this approach  in Sect.~\ref{sec:non_adiabatic}.

\subsection{Cancellation frequency \label{sec:cancellation_frequency}}

In the solar case, \citetalias{Berthomieu1990} already emphasised the existence of a cancellation frequency $\nu_{c,\ell}$ for each degree $\ell$ , where, notwithstanding the mode r.m.s velocity, no luminosity fluctuation will be observable. 
From Eq.~(\ref{eq:deltaL_L}), it is straightforward to check that this cancellation phenomenon happens when 
\begin{equation}
\overline{\omega}^4 + \left[ 4 + \frac{1}{4\nabla_\mathrm{ad} b_\ell} (c_\ell - 2 b_\ell) \right] \overline{\omega}^2 - \ell (\ell+1) = 0\; .
\end{equation}
Real solutions $\overline{\omega}_c$ for this equation verify 
\begin{equation}
\label{eq:cancellation_frequency}
\begin{aligned}
\overline{\omega}^2_c (\ell) &= \frac{1}{2} \Bigg\{ - 4 - \frac{1}{4\nabla_\mathrm{ad} b_\ell} (c_\ell - 2 b_\ell) \\
&+ \left[ \left( 4 + \frac{1}{4\nabla_\mathrm{ad} b_\ell} (c_\ell - 2 b_\ell) \right)^2 + 4 \ell(\ell+1) \right]^{1/2}
\Bigg\} \; .
\end{aligned}
\end{equation}

\begin{figure}[ht!]
    \centering
    \includegraphics[width=0.48\textwidth]{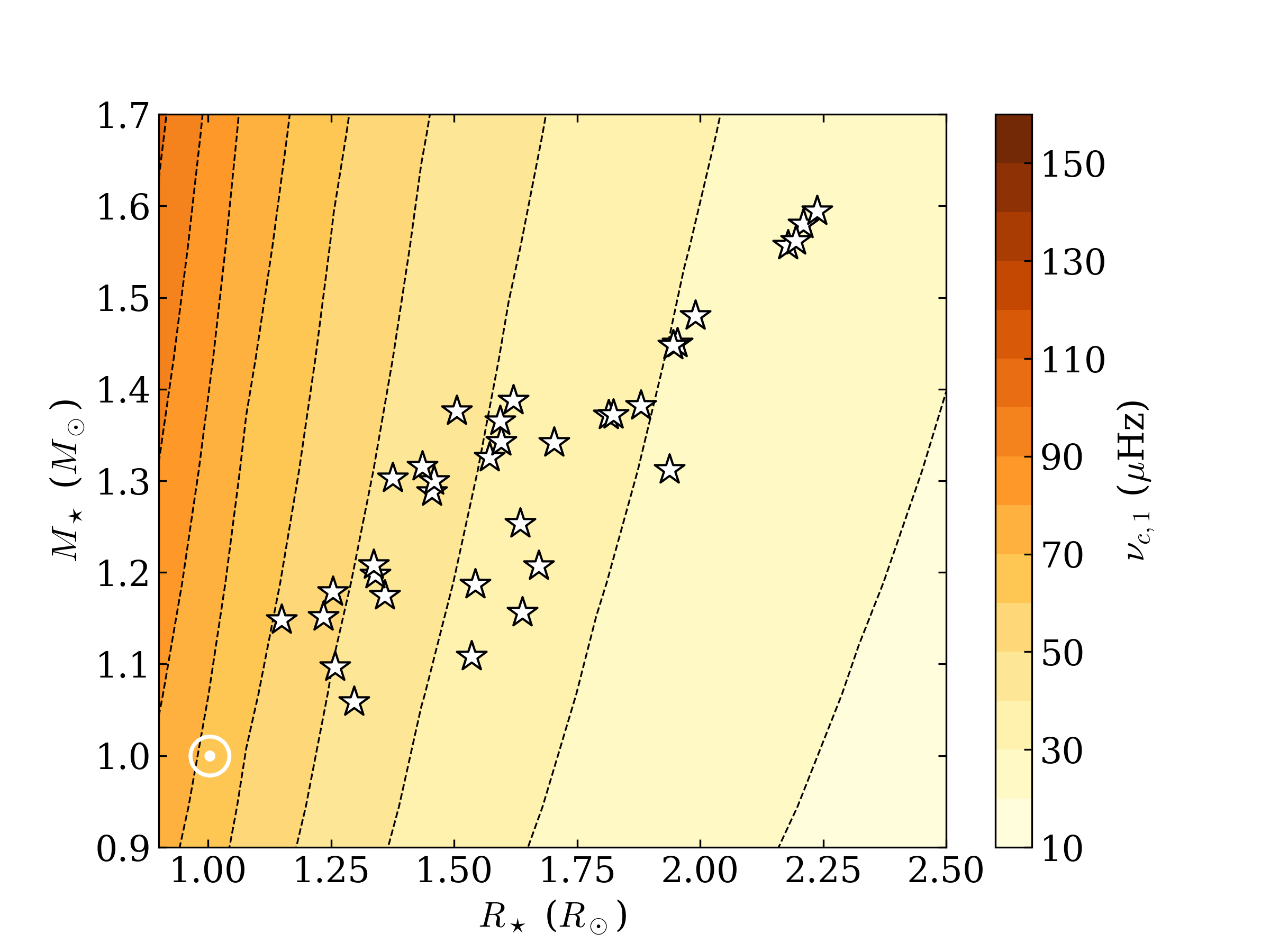}
    \includegraphics[width=0.48\textwidth]{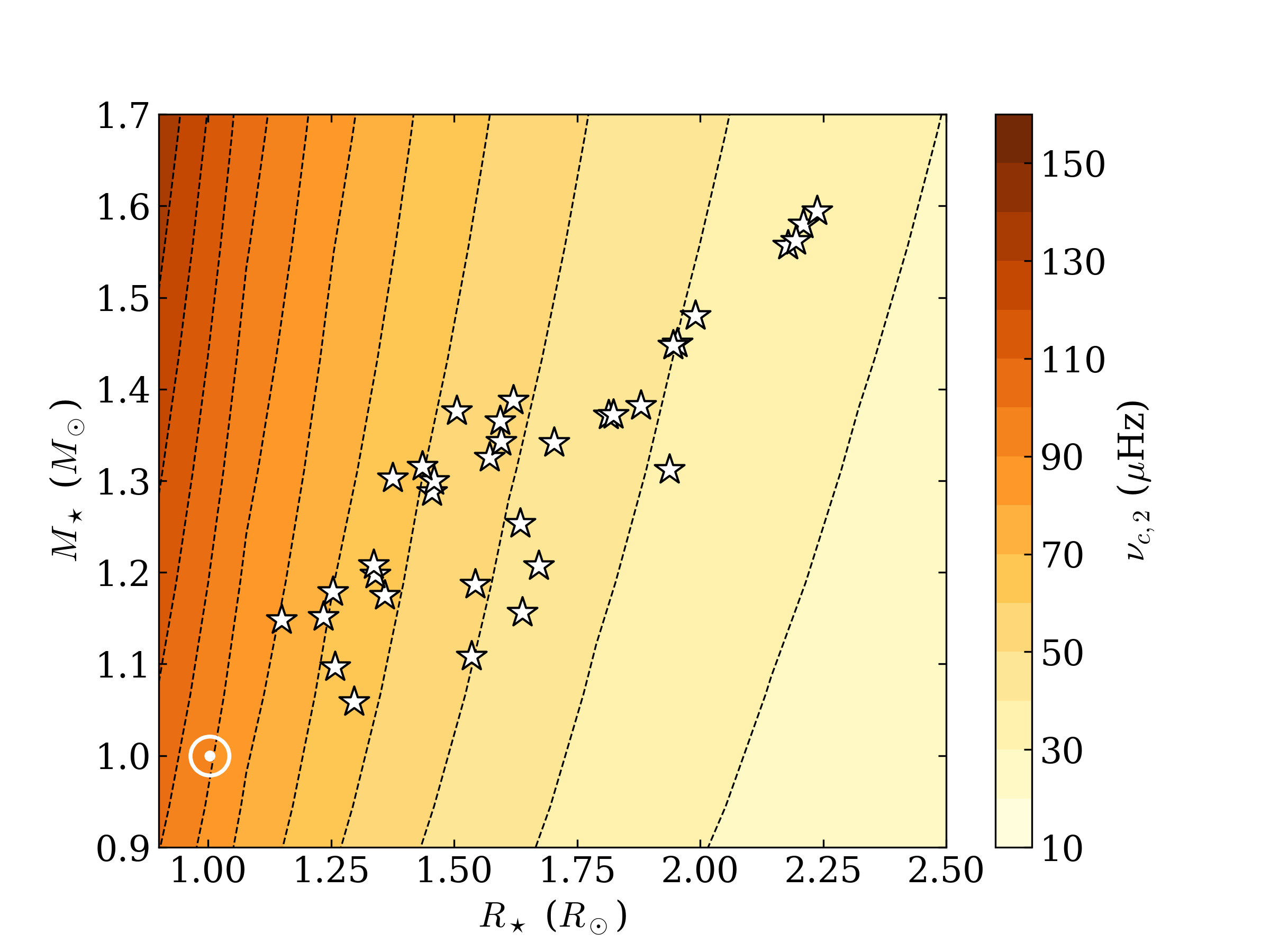}
    \includegraphics[width=0.48\textwidth]{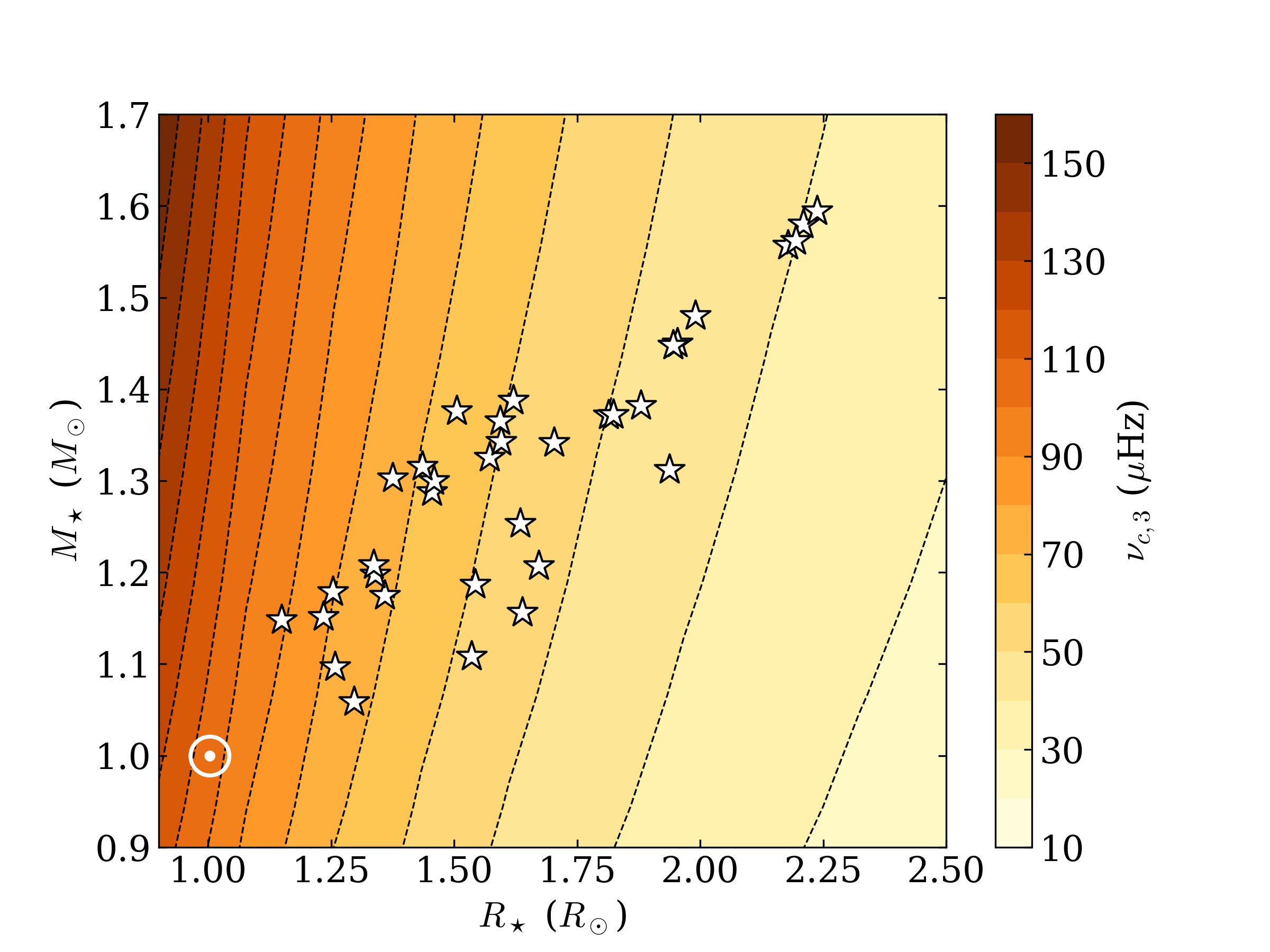}
    \caption{Cancellation frequency $\nu_{c,1}$ (\textit{top}), $\nu_{c,2}$ (\textit{middle}), and $\nu_{c,3}$ (\textit{bottom}) in the adiabatic case as a function of radius $R_\star$ and mass $M_\star$. The stars of our sample are shown by white stars, and the Sun is represented for comparison with its usual symbol, $\odot$ in white.
    }
    \label{fig:nu_cancellation_grid}
\end{figure}

It is interesting to note that, from the relation between $\overline{\omega}$ and $\nu$, the cancellation frequency $\nu_{c,\ell}$ scales like $M_\star / R_\star^3$. 
We illustrate this in Fig.~\ref{fig:nu_cancellation_grid} by showing the evolution of $\nu_{c,\ell}$ with $R_\star$ and $M_\star$ for $\ell=1$, 2, and 3. We can see that, given the extent of our sample, $\nu_{c,\ell}$ may significantly differ between the targets we consider. Contrary to the solar case, we can also note that it remains important for late F-type stars to take into account the cancellation effect on $\ell=2$ and 3 modes when studying frequency regions below 100~$\mu$Hz. 

The main issue with this $\nu_{c,\ell}$ quantity is that it strongly relies on the adiabatic assumption that we make above. Non-adiabatic surface effects can indeed have a significant influence on the value of the cancellation frequency, which prevent us from using this value as an accurate tool to select regions that should be privileged when searching for g-mode signatures. 

\subsection{Non-adiabatic effects \label{sec:non_adiabatic}}

To obtain Eq.~(\ref{eq:deltaL_L_rms}) and \ref{eq:cancellation_frequency}, we neglected the non-adiabatic surface effects that should be included in the energy equation of the stellar oscillation and we considered the simple case of adiabatic oscillations. This approach presents a substantial advantage in that it connects the mean intensity fluctuation to the surface displacement through an analytical relation, which circumvents the need to numerically solve the oscillation equations \citepalias{Berthomieu1990}. From the computations performed by \citetalias{Berthomieu1990} for the non-adiabatic solar case, it appears that $\delta L_\star / L_\star$ may vary by a factor with an order of magnitude of unity depending on the assumptions chosen when performing the non-adiabatic computations.   
It is therefore expected that variations of the same order should occur in the late-F-type star case.
The interplay between oscillations and convection, which is neglected by \citetalias{Berthomieu1990}, might also be of importance in order to obtain an accurate result \citep[e.g.][]{Bunting2019}. However, we find that the adiabatic approximation is enough to provide us with an idea of the mode-velocity order of magnitude necessary to induce a detectable luminosity fluctuation. 
This quantity is strongly model dependent, but our goal in this work is not to provide the most accurate value of the cancellation frequency, but to illustrate that this is an effect that cannot be ignored. It has to be emphasised that g-mode detection in solar-type stars could provide additional constraints on mode surface effects and improve the modelling of p modes \citep[e.g.][]{Kjeldsen2008,Ball2017,Houdek2017}.  

\subsection{Velocity levels}

We now turn to the computation of the velocity levels required for adiabatic oscillation modes to induce a luminosity fluctuation detectable among the background stellar signal.
To simplify our analysis, we consider the case $\theta_0 = 90^o$, where the observer line of sight belongs to the stellar equatorial plane.
In the presence of rotation, assuming that the mode lifetime is significantly larger than the stellar rotation period, a power peak detected in the PSD should correspond to an individual mode component of degree $\ell$ and azimuthal number $m$. Following \citetalias{Berthomieu1990}, we consider the $m=0$ component for even $\ell$ and the $m=1$ component for odd $\ell$. We remind the reader that, for $\theta_0 = 90^o$, components with odd $\ell - m$ are not visible for disc-integrated observations.  
We also underline that our analysis does not lack generality in the sense that, for any $\theta_0$, it is possible for a given degree $\ell$ to select the $m$-component with largest amplitude by considering the profile of the $P_\ell^m$ polynomial as a function of $\cos \theta_0$.  
However, doing this would add unnecessary complexity to the analysis we present here.  

\begin{figure}[ht!]
    \centering
    \includegraphics[width=.48\textwidth]{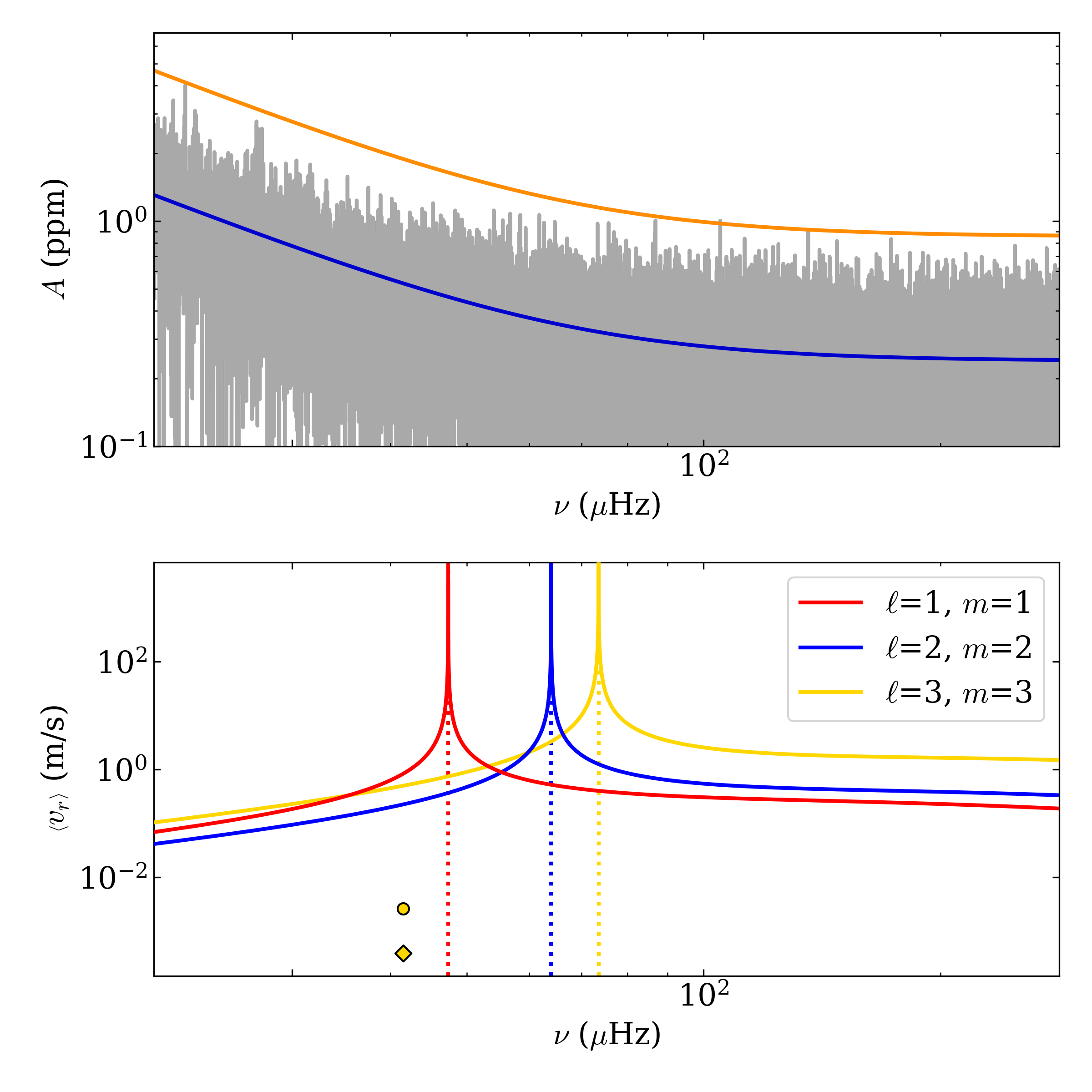}
    \caption{
    Example of stellar background and corresponding mode velocity levels required to induce a detectable luminosity fluctuation.
    \textit{Top:} Power spectrum (\textit{grey}) of KIC~3733735.
    The solid blue line corresponds to the $B$ profile. The corresponding significance level computed with Eq.~(\ref{eq:appourchaux2000}) for $p_\mathrm{det} = 0.1$ is shown by the solid orange line. The power spectrum was rescaled from a power spectral density to an amplitude in ppm to simplify the comparison with the luminosity fluctuation level.
    \textit{Bottom:} $\left<v_r\right>$ levels necessary to produce a detectable luminosity fluctuation, assuming the background profiles computed for KIC~3733735, for $\ell=1$, $m=1$ (\textit{red}), $\ell=2$, $m=0$ (\textit{blue}), and $\ell=3$, $m=1$ (\textit{yellow}) modes.  
    In each case, the position of the cancellation frequency $\nu_{c,\ell}$ is emphasised by a vertical dotted line of the proper colour. The two yellow symbols correspond to the $\left<v_r\right>$ obtained for the $\ell=3$ mode with highest amplitude in the 3D hydrodynamical simulations from \citetalias{Breton2022simuFstars}, that is, at the top of the radiative zone (\textit{circle}) and close to the top of the simulation domain (\textit{diamond}), respectively.
    }
    \label{fig:rms_velocity_level}
\end{figure}

By renormalising the PSD, we obtain the background amplitude level in ppm (and not in ppm$^2$/$\mu$Hz as most commonly used in seismology of solar-type stars and shown for example in Fig.~\ref{fig:fit_background_parametric}). This amplitude level therefore provides us with a measurement of $\left<\delta L_\star / L_\star \right>$ as a function of frequency, which we can use to deduce, through Eq.~(\ref{eq:deltaL_L_rms}), the $\left<v_r\right>$ amplitude necessary for an oscillation mode to induce a detectable luminosity fluctuation at a given frequency ---accounting for the background signal--- provided that we are able to estimate the detection level, which we compute here through Eq.~(\ref{eq:appourchaux2000}).
We therefore show in Fig.~\ref{fig:rms_velocity_level} the peak significance level $s_\mathrm{det}$ computed with Eq.~(\ref{eq:appourchaux2000}), as well as the corresponding $\left<v_r\right>$ values necessary for a mode of given frequency to induce a luminosity fluctuation corresponding to this significance level in the case of KIC~3733735. 
We find very similar behaviours for the other stars of our sample. We obtain this $\left<v_r\right>$ profile from Eq.~(\ref{eq:deltaL_L_rms}). 
Because of visibility effects related to the integrals defined in Eq.~(\ref{eq:b_l_c_l}), $\ell=3$ modes require large amplitude in order to be detected in disc-integrated observations.
For any given $\ell$, in the neighbourhood of the vertical asymptote of the cancellation frequency, even a low-amplitude luminosity fluctuation requires an unrealistically large g-mode r.m.s velocity to be detectable.

Above the cancellation frequency, $\ell=1$ modes are those that require the least amplitude to be detectable in photometric data, justifying our attempt to identify the patterns shown in Sect.~\ref{sec:kic3733735} and \ref{sec:mixed_modes} with $\ell=1$ frequencies computed with GYRE.
It is interesting to note that below the cancellation frequency, the photometric significance level corresponds to lower $\left<v_r\right>$ levels ---that is, of around $\sim0.1$~m/s--- for $\ell=2$ modes than for $\ell=1$ modes, for which we have $\left<v_r\right> \sim 0.2$~m/s. 
This can be explained through the role the non-radial term $\ell (\ell+1) / \overline{\omega}^2$  plays in the relation between velocity and luminosity fluctuation. Indeed, for low-frequency modes, $\overline{\omega}$ is smaller than unity. Therefore, this non-radial contribution dominates asymptotically and, for a given $\left < v_r \right >$, the resultant local luminosity fluctuation given in Eq.~(\ref{delta_L_L_non_integrated}) increases with $\ell$. For $\ell \geq 3$, this effect is counterbalanced by the integration on the disc and the small ratios $b_\ell / b_1$. For $\ell=2$, in the case of KIC~3733735, we have $b_2 / b_1 \sim 0.455,$ while the ratio between the non-radial terms is 3. The $\sqrt{(2\ell+1)(\ell - m)!/(\ell + m)!} |P_\ell^m(0)|$ prefactors ratio should also be considered, and is $\sqrt{5}/2$ in this case. Finally, we expect to have, asymptotically, $\left<v_r\right>_{\ell=1} / \left<v_r\right>_{\ell=2} \sim 1.53$, which is consistent with what we obtain in Fig~\ref{fig:rms_velocity_level}.
We remind that in a scenario assuming equipartition of energy, the $\left<v_r\right>$ level would be lower for $\ell=2$ \citepalias[see][]{Berthomieu1990}.
However, this is particularly interesting from the perspective of forced tidal oscillations, where equipartition is not respected and $\ell = 2$ modes should be excited more efficiently than $\ell=1$ modes. For comparison, using the scaling relation provided by \citet{Kjeldsen1995}, the p modes in a star with $M_\star = 1.3$~$\rm M_\odot$ and luminosity $L_\star = 3.8$~$\rm L_\odot$ are expected to have an average amplitude of around 0.7~m/s.

Below the cancellation frequency, a mode with a given $\left<v_r\right>$ produces a larger radial displacement than its higher-frequency counterparts of the same $\left<v_r\right>$, and therefore induces a larger luminosity fluctuation. 
For example, a $\ell=2$ mode with frequency $\nu = 25$~$\mu$Hz would only need a $\num{6e-2}$~m/s velocity to produce a detectable fluctuation in the power spectrum. 
The $\left<v_r\right>$ threshold above which a detectable signal is produced is therefore lower for low-frequency modes. In particular, theoretical considerations from \citet{Augustson2020} and 3D simulations performed in \citetalias{Breton2022simuFstars} suggest that low-frequency modes are more efficiently excited for a fast rotating star, and may also possibly tunnel more efficiently through the convective envelope \citep{Mathis2023Tunneling}. However, considering the highest-amplitude $\ell=3$ mode (we remind the reader that no signature was detected at the top of the simulation domain for $\ell=1$ and 2 modes) reported by \citetalias{Breton2022simuFstars}, we note that, considering the background level obtained from \textit{Kepler} observations, the mode amplitudes obtained from simulations are two to three orders of magnitude below the level necessary to induce a detectable luminosity fluctuation. It is important to remember that, due to numerical constraints, the radiative diffusivity responsible for the mode damping \citep{Zahn1997,Alvan2014} is significantly larger in 3D numerical simulations than in actual stellar interiors \citep[e.g.][]{Garaud2021}. Three-dimensional simulations are nevertheless powerful tools for understanding the non-linear behaviour of oscillation modes in stellar interiors. In particular, they allow us to study the transfer function from the convection spectrum to the wave spectrum. Comparing their predictions to actual observations would therefore allow us to reach a better understanding of the efficiency of the complex interactions between convective motions and internal waves.  

While it is unlikely that stellar fluid regimes will soon be reached in global spherical simulations of the dynamics of stars, there is potential to get closer to these regimes. A first option to reduce the computing cost (and therefore increase the turbulence of the simulations) is to work in a 2D configuration with polar coordinates \citep[e.g.][]{Rogers2013,LeSaux2022} or to study more localised phenomena using 2D or 3D cartesian boxes 
\citep[e.g.][]{Kiraga2005,Lecoanet2015}. However, choosing such options means losing the spherical geometry of the problem and the straightforward comparison that it allows with mode properties computed by linear oscillation codes. A promising perspective is offered by adaptive mesh refinement (AMR) grids \citep{Berger1984,Berger1989}, which allow us to adapt the grid resolution in turbulent areas in each integration step while keeping large mesh grids in quiescent regions. Effort is currently being dedicated to developing MHD AMR codes designed for the study of stellar dynamics \citep{Delorme2022}.

\begin{figure*}[ht!]
    \centering
    \includegraphics[width=\textwidth]{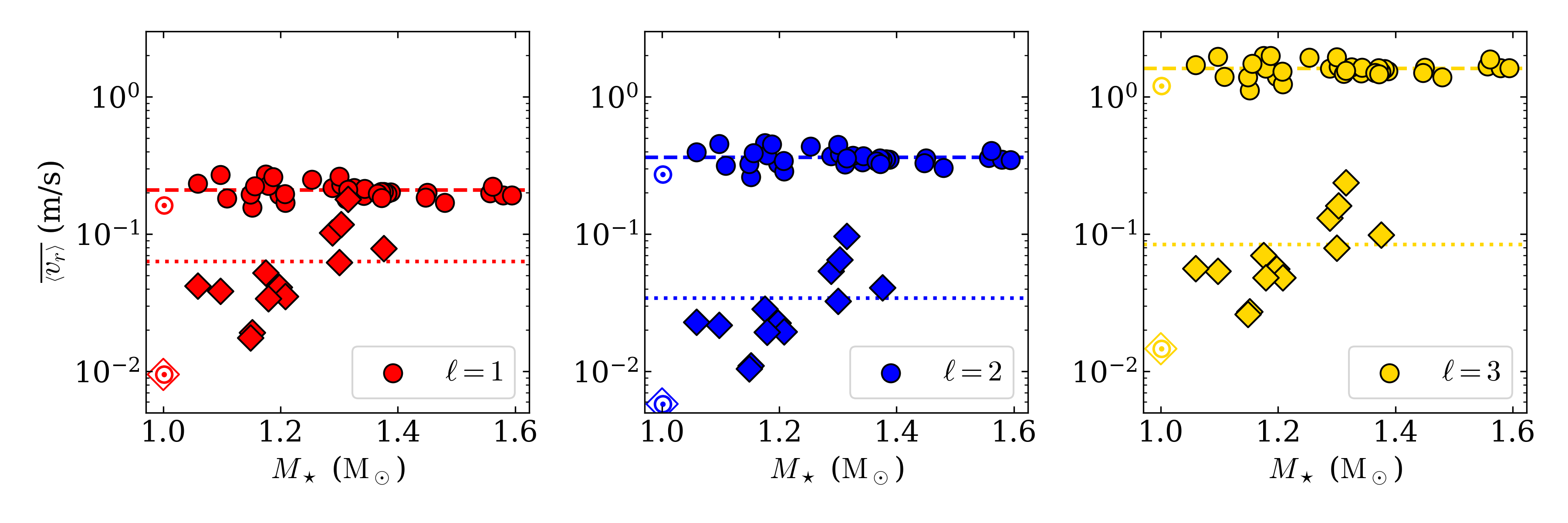}
    \caption{
    Frequency-averaged velocity level $\overline{\left<v_r\right>}$ as a function of $M_\star$ measured above 125~$\mu$Hz (\textit{circles}, $\overline{\left<v_r\right>}_+$) and below 30~$\mu$Hz (\textit{diamonds}, $\overline{\left<v_r\right>}_-$) for the stars in our sample and for $\ell=1$ (\textit{left}), $\ell=2$ (\textit{centre}), and $\ell=3$ (\textit{right}) modes. 
    To avoid biases in the comparison, we do not consider $\overline{\left<v_r\right>}_-$ for stars with $\nu_{c,1} < 40$~$\mu$Hz.
    In each panel, the mean values of the $\overline{\left<v_r\right>}_+$ and $\overline{\left<v_r\right>}_-$ distributions are shown by the dashed and dotted lines, respectively.
    In each panel, the mean velocity levels computed with the VIRGO/SPM solar time series $\overline{\left<v_r\right>}$ are shown using the usual solar symbol, $\odot$. The solar symbol enclosed by a square corresponds to $\overline{\left<v_r\right>}_+$, while the other corresponds to $\overline{\left<v_r\right>}_-$.
    }
    \label{fig:mean_velocity_level}
\end{figure*}

Finally, in order to provide a summary diagnostic for our analysis, we represent in Fig.~\ref{fig:mean_velocity_level} the frequency-averaged velocity level $\overline{\left<v_r\right>}$ above 125~$\mu$Hz and below 30~$\mu$Hz as a function of stellar mass $M_\star$. We denote these two values $\overline{\left<v_r\right>}_+$ and $\overline{\left<v_r\right>}_-$, respectively.
The bounds on which the averages are computed are chosen to avoid the bias that the asymptotic behaviour around the cancellation frequency would introduce. It should be noted that, for the same reason, we do not compute $\overline{\left<v_r\right>}_-$ for stars where $\nu_{c,1} < 40$~$\mu$Hz.

Below 30~$\mu$Hz, the mean value for $\overline{\left<v_r\right>_-}$ is \num{6.3e-2}~m/s for $\ell=1$ modes, \num{3.4e-2}~m/s for $\ell=2$ modes, and \num{8.4e-2}~m/s for $\ell=3$.
Above 125~$\mu$Hz, the mean value for $\overline{\left<v_r\right>_+}$ is \num{2.1e-1}~m/s for $\ell=1$ modes, \num{3.6e-1}~m/s for $\ell=2$ modes, and \num{1.6}~m/s for $\ell=3$. A clear trend is visible in this case, as the velocity level required to induce a detectable fluctuation increases with the mass. These values are summarised in Table~\ref{tab:mean_v_r_values}. As already underlined with the specific case of KIC~3733735, below the cancellation frequency, a lower value of $\left<v_r\right>$ is required to induce a detectable luminosity fluctuation, which also puts more stringent constraints on the upper limit of the actual surface g mode velocity for this ranges of frequency. 
In order to provide a comparison with the solar case, we perform the same analysis with a four-year time series from the Sunphotometers of the Variability of Solar Irradiance and Gravity Oscillations \citep[VIRGO/SPM,][see Appendix~\ref{appendix:sun} for more details]{Frohlich1995}. 
Using the formalism from Eq.~(\ref{eq:deltaL_L_rms}) once again, we find that the velocity level necessary to generate a detectable luminosity fluctuation at the surface of the Sun is below the mean values measured for the F-type sample, both above and below the cancellation frequency. 
This is expected as the granulation power level in the PSD increases as $\nu_\mathrm{max}$ decreases \citep[see e.g.][]{Kallinger2014}, that is, on the MS, roughly as $M_\star$ increases. 

\begin{table}
    \centering
    \caption{Mean $\overline{\left<v_r\right>_-}$ and $\overline{\left<v_r\right>_+}$ as a function of $\ell$.}
    \begin{tabular}{cccc}
    \hline\hline
    $\ell$ & 1 & 2 & 3 \\
    \hline
    $\overline{\left<v_r\right>_-}$ (m/s) & \num{6.3e-2} & \num{3.4e-2} & \num{8.4e-2} \\
    $\overline{\left<v_r\right>_+}$ (m/s) & \num{2.1e-1} & \num{3.6e-1} & \num{1.6} \\
    \end{tabular}
    \label{tab:mean_v_r_values}
\end{table}

As a final comment, we underline that the comparison we provide here should be interpreted as being in favour of g-mode detectability in late F-type stars. Indeed, the $\left< v_r \right>$ thresholds that we obtain here for the Sun remains of the same order of magnitude as the ones obtained for F-type stars. From analytical predictions  \citet[e.g.][]{Pincon2016}, we nevertheless expect  the power injection from pummeling and turbulent convection to the modes to be several orders of magnitude larger in late F-type stars than in the Sun. This last point is observed in 3D simulations, and was discussed in Sect.~6 of \citetalias{Breton2022simuFstars}.  

\section{Conclusion \label{section:conclusion}}

In this work, we attempted to detect stochastically excited g modes in MS solar-type stars. For this purpose, we selected and analysed the most promising sample of stars for which four-year \textit{Kepler} time series are available. These stars are late F-type stars with thin convective envelopes and fast convective flows. They all exhibit the signature of stochastically excited p modes in the \textit{Kepler} short-cadence time series. Having underlined the difficulties in accurately measuring the level of background noise (dominated by stellar granulation in this frequency range) in the presence of rotational power modulation, we performed a statistical analysis to detect significant peaks and provide a list of stars where individual peaks are detected with a 90~\% probability of not being the product of noise, a threshold that was used in an attempt to detect g mode signatures in solar observations. Visually inspecting the S/N spectra of the stars with detected individual peaks, we selected patterns of peaks with a low probability of being the product of noise (below 0.5~\%) and compared the configuration of these patterns to the GYRE-predicted frequencies of $\ell=1$ modes using a reference stellar model for each star where we observed such a pattern.
We first discuss the case of KIC~3733735, where the detected signal is compatible with the presence of the $n=1$, $\ell=1$ p mode, and, at a lower frequency, a pattern of low-order non-asymptotic $\ell=1$ g modes. 
We underline that the differences between the model and the observations could be explained by the fact that g-mode period spacing is strongly sensitive to the extent of the convective core of the star and therefore to the physical modelling of core overshooting and chemical mixing. This highlights the potential to reach a better calibration of these processes in stellar evolution models  in the future.
Considering the possibility of a coupling between the p- and g- mode resonant cavities, we then discuss the case of three additional targets, KIC~6679371, KIC~7103006, and KIC~9206432, using a GYRE-computed frequency again for comparison. We use our reference models to show that non-asymptotic modes laying in this range of frequency are indeed expected to have mixed characters. Their properties should therefore be strongly affected by the coupling between the p- and g-mode resonant cavities.  
Characterising mixed modes in MS solar-type stars would therefore allow us to probe yet-unexplored layers  in MS solar-type stars and to bring strong structural and evolutionary constraints in this mass range. In particular, we underline the fact that the properties {of non-asymptotic} pure g and mixed modes should be studied in more detail in the case of late F-type stars, as it appears that their properties are strongly influenced by the stratification profile at the convective core interface. We also note that GYRE predicts the existence of non-asymptotic pure g modes (i.e. modes without a node in the p-mode resonant cavity) with low normalised inertia and a significant proportion of their energy concentrated in the upper layer of the star, suggesting that these mode properties are still affected by the vicinity of the p-mode resonant cavity.

Considering the detection levels obtained from our statistical analysis, we use the formalism from \citetalias{Berthomieu1990} to compute an equivalent r.m.s velocity level necessary to trigger a detectable luminosity fluctuation as a function of frequency. 
The objective of this analysis is to derive an upper threshold for g-mode velocity in late F-type stars as well as to obtain an estimate of the mode velocity in the frequency windows where we suggest that the signal we detect is related to mixed modes.
We emphasise the existence of a photometric cancellation frequency. Combined with the detection of oscillation modes in the same frequency window, this could help to characterise non-adiabatic effects in the convective envelope.
Using an adiabatic approximation, we show that g-mode oscillations in late F-type stars should have a maximal amplitude of about a few tens of cm/s for g modes with the highest frequency, and of a few cm/s for low-frequency g modes. The candidate modes we identify in this work should therefore have surface velocities of $\sim$10 cm/s, a result that can be used in order to bring constraints on the efficiency of power injection from convection to stochastic oscillation modes.

While these results are promising, further analysis of a larger sample is necessary in order to confirm their validity. Fortunately, the upcoming PLAnetary Transits and Oscillations of stars \citep[PLATO,][]{Rauer2014} mission, which is designed to characterise oscillations in MS solar-type stars, will enable us to increase our sample size by at least one order of magnitude.

\begin{acknowledgements}
The authors thank the anonymous referee for constructive suggestions and comments that allowed significant improvement of the manuscript. 
They also thank A.~Jiménez for providing the VIRGO timeseries used for the solar comparison and S.~Gebruers for providing the machine-readable table used to represent the $\gamma$ Dor sample on Fig.~\ref{fig:teff_logg_diagram}. They also acknowledge L.~Amard, L.~Bugnet, S.~Sulis, and L.~Bigot for fruitful discussions. 
S.N.B, HD, R.A.G, St.M, and A.S.B acknowledge support from PLATO and GOLF CNES grants. 
S.N.B, and A.S.B acknowledge support from ERC Whole Sun Synergy grant \#810218. 
S.N.B acknowledges support from PLATO ASI-INAF agreement n.~2015-019-R.1-2018.
Sa.M.~acknowledges support from the Spanish Ministry of Science and Innovation (MICINN) with the Ram\'on y Cajal fellowship no.~RYC-2015-17697, grant no.~PID2019-107187GB-I00 and PID2019-107061GB-C66, and through AEI under the Severo Ochoa Centres of Excellence Programme 2020--2023 (CEX2019-000920-S).
A.R.G.S. acknowledges the support by FCT through national funds and by FEDER through COMPETE2020 by these grants: UIDB/04434/2020 \& UIDP/04434/2020. A.R.G.S. is supported by FCT through the work contract No. 2020.02480.CEECIND/CP1631/CT0001.
This paper includes data collected by the \textit{Kepler} mission, and obtained from the MAST data archive at the Space Telescope Science Institute (STScI). Funding for the \textit{Kepler} mission is provided by the NASA Science Mission Directorate. STScI is operated by the Association of Universities for Research in Astronomy, Inc., under NASA contract NAS 5–26555.
\\
\textit{Software:} 
MESA \citep{Paxton2011, Paxton2013, Paxton2015, Paxton2018, Paxton2019}, 
GYRE \citep{Townsend2013,Townsend2018,Goldstein2020}, 
\texttt{numpy} \citep{harris2020array}, \texttt{matplotlib} \citep{Hunter:2007}, \texttt{scipy} \citep{2020SciPy-NMeth}, \texttt{emcee} \citep{Foreman-Mackey2013}, \texttt{corner} \citep{corner}, \texttt{astropy } \citep{astropy:2022}, \texttt{apollinaire} \citep{Breton2022apollinaire}.
\end{acknowledgements}

\bibliographystyle{aa} 
\bibliography{biblio} 

\appendix

\section{VIRGO/SPM solar photometric time series \label{appendix:sun}}

\begin{figure}[ht!]
    \centering
    \includegraphics[width=0.48\textwidth]{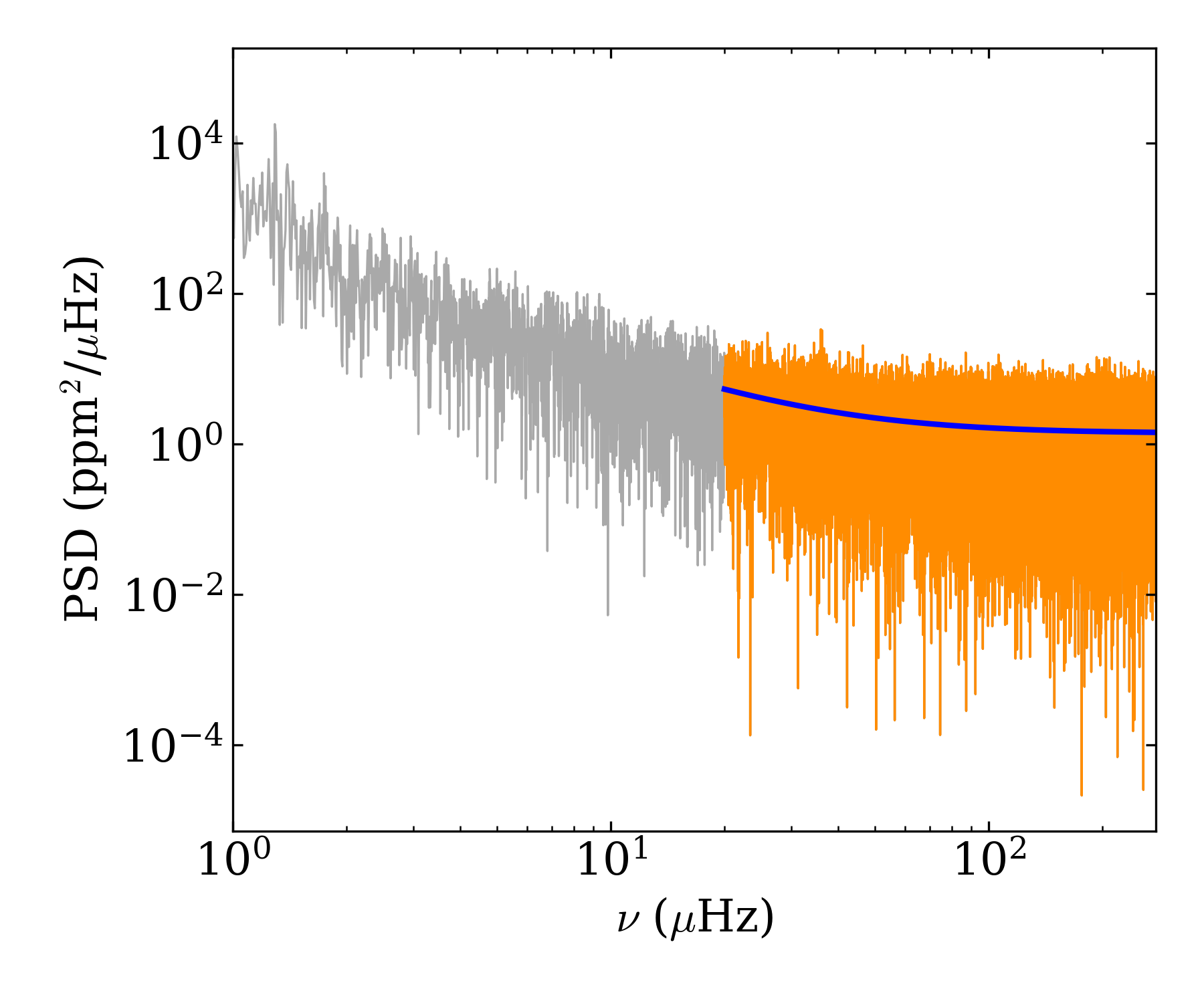}
    \caption{Background model for the Sun using the combination of the green and red channels from VIRGO/SPM. Only the frequency bins of the PSD shown in dark orange are considered to obtain the background. The low-frequency area of the PSD (in grey) is not considered.}
    \label{fig:sun_psd_background}
\end{figure}

In order to provide a comparison with the late F-type stars data acquired by \textit{Kepler}, we consider the observations from VIRGO/SPM. 
To match the observing bandwidth of the \textit{Kepler} satellite, we use the combination of the time series acquired by the red and green channels of VIRGO/SPM \citep{Basri2010,Salabert2017}. 
We consider a four-year time series during a minimum of the solar cycle between 2006 and 2010. The parametric and non-parametric background profiles are then obtained following the procedure described in Sect.~\ref{sec:ensemble_analysis}.
We show in Fig.~\ref{fig:sun_psd_background} the PSD computed from the VIRGO/SPM time series and the corresponding background profile we compute with the 20~$\mu$Hz low-frequency cutoff.

\section{Stellar modelling \label{appendix:modelling}}

The reference models for KIC~3733735, KIC~6679371, KIC~7103006, and KIC~9206432 used in this work were computed using the IACgrid, which is described in detail in \citet{Gonzalez-Cuesta2023}. The grid was generated with MESA version 15140 and covers the mass range from 0.8 to 1.5~$M_\odot$. Apart from the OPAL opacities \citep{Iglesias1996} and the GS98 metallicity mixtures \citep{Grevesse1998}, standard MESA input physics are used. The models therefore do not include core overshooting.
Input observables used for the modelling include spectroscopic $T_\mathrm{eff}$, $\log g$, and $\rm [Fe/H]$ (see Table~\ref{tab:fstar_selection}), \textit{Gaia} luminosity computed by \citet{Berger2020}, global asteroseismic parameters $\nu_\mathrm{max}$ (see Table~\ref{tab:param_sismo}), and p-mode frequencies, for which we use the values provided by \citet{Lund2017} except for KIC~3733735 for which no published table of frequency is available. For this star, we fit the p mode frequency in the PSD of the short-cadence data with the \texttt{apollinaire} module \citep{Breton2022apollinaire}. In Table~\ref{tab:mode_frequency_3733735}, we provide the orders $n$, the degree $\ell$, and the frequency $\nu$ of the modes used for the modelling. 
Observables are fitted on the grid following a $\chi^2$ minimisation as described in \citet{Perez-Hernandez2016,Perez-Hernandez2019}. The p-mode frequencies are accounted for with a surface term correction following the procedure detailed in \citet{Perez-Hernandez2019}. We underline that the model we obtain for KIC~6679371 is located at the edge of the modelling grid, and that the uncertainties we provide for it in Table~\ref{tab:fstar_selection} are probably underestimated. Nevertheless, the reference model that we consider is compatible with the input observables.

\begin{table}[ht!]
\centering
\caption{Mode frequencies fitted with \texttt{apollinaire} for KIC~3733735.}
\label{tab:mode_frequency_3733735}
\begin{tabular}{ccr}
\hline \hline
$n$ & $\ell$ & $\nu$ ($\mu$Hz) \\
\hline
13 &  2 & $1422.27 \pm 1.55$ \\
14 &  0 & $1428.10 \pm 1.63$ \\
14 &  1 & $1473.52 \pm 0.59$ \\
14 &  2 & $1512.21 \pm 1.48$ \\
15 &  0 & $1517.03 \pm 1.15$ \\
15 &  1 & $1562.33 \pm 0.60$ \\
15 &  2 & $1601.27 \pm 1.96$ \\
16 &  0 & $1606.88 \pm 0.81$ \\
16 &  1 & $1652.60 \pm 0.50$ \\
16 &  2 & $1693.21 \pm 1.31$ \\
17 &  0 & $1698.77 \pm 0.67$ \\
17 &  1 & $1745.19 \pm 0.56$ \\
17 &  2 & $1786.41 \pm 1.18$ \\
18 &  0 & $1792.29 \pm 0.68$ \\
18 &  1 & $1839.02 \pm 0.49$ \\
18 &  2 & $1878.04 \pm 0.75$ \\
19 &  0 & $1886.43 \pm 0.70$ \\
19 &  1 & $1933.03 \pm 0.71$ \\
19 &  2 & $1975.22 \pm 2.37$ \\
20 &  0 & $1977.78 \pm 1.18$ \\
20 &  1 & $2024.30 \pm 0.57$ \\
20 &  2 & $2065.47 \pm 1.64$ \\
21 &  0 & $2070.34 \pm 1.37$ \\
21 &  1 & $2116.11 \pm 0.48$ \\
21 &  2 & $2159.85 \pm 2.08$ \\
22 &  0 & $2161.39 \pm 0.98$ \\
22 &  1 & $2208.94 \pm 0.63$ \\
23 &  0 & $2253.18 \pm 0.97$ \\
23 &  1 & $2299.99 \pm 0.58$ \\
24 &  0 & $2344.64 \pm 1.25$ \\
24 &  1 & $2392.03 \pm 0.68$ \\
\end{tabular}
\end{table}

\end{document}